\documentclass[12pt,preprint]{aastex}

\shorttitle{}
\shortauthors{}

\begin{document}

\title{Formation of the First Stars by Accretion}
\author{Kazuyuki Omukai \altaffilmark{1}
and Francesco Palla \altaffilmark{2}}
\altaffiltext{1}{Division of Theoretical Astrophysics, 
National Astronomical Observatory, Mitaka, Tokyo 181-8588, Japan}
\altaffiltext{2}{INAF--Osservatorio Astrofisico di Arcetri, 
Largo E. Fermi 5, 50125 Firenze, Italy}
\email{omukai@th.nao.ac.jp; palla@arcetri.astro.it}
\begin{abstract}
The process of star formation from metal-free gas is investigated 
by following the evolution of accreting protostars with emphasis on
the properties of massive objects. The main aim is to establish
the physical processes that determine the upper mass limit of the
first stars. Although the consensus is that massive stars were commonly
formed in the first cosmic structures, our calculations show that their 
actual formation depends sensitively on the mass accretion rate and
its time variation. Even in the rather idealized case in which star
formation is mainly determined by $\dot{M}_{\rm acc}$, the characteristic
mass scale of the first stars is rather uncertain. We find that there is
a critical mass accretion rate $\dot{M}_{\rm crit}\simeq 4\times 10^{-3} 
M_\sun {\rm yr}^{-1}$ that separates solutions with 
$\dot{M}_{\rm acc}< \dot{M}_{\rm crit}$ in which objects with
mass $\gg 100 M_\sun$ can form, provided there is sufficient matter
in the parent clouds, from others ($\dot{M}_{\rm acc}> \dot{M}_{\rm crit}$) 
where the maximum mass limit decreases as $\dot{M}_{\rm acc}$ increases. 
In the latter case, the protostellar
luminosity reaches the Eddington limit before the onset of hydrogen 
burning at the center via the CN-cycle. This phase is followed by a
rapid and dramatic expansion of the radius, possibly leading to reversal 
of the accretion flow when the stellar mass is about 100~$M_{\sun}$.

Under a realistic time dependent accretion rate that starts at high values 
($\sim 10^{-2} M_\sun$ yr$^{-1}$) and decreases rapidly in 
the high mass regime ($M_\ast \ga 90 M_\sun$), the evolution follows 
the case of $\dot{M}_{\rm acc}<\dot{M}_{\rm crit}$ and accretion can continue
unimpeded by radiation forces. Thus, the maximum mass is set by consideration
of stellar lifetimes rather than by protostellar evolution. In this case,
the upper limit can be as high as $\sim 600 M_\sun$.

We consider also the sensitivity of the results to the presence of 
heavy elements with abundances in the range
$Z=5\times 10^{-5} Z_{\sun}$ to $5 \times 10^{-3} Z_{\sun}$.  The main
evolutionary features of protostars are similar to
those of metal-free objects, except that the value of $\dot{M}_{\rm crit}$
increases for metal-enriched protostars.  Since the accretion rate is
lower in a slightly polluted environment, the condition $\dot{M}_{\rm acc} <
\dot{M}_{\rm crit}$ is expected to be more easily met.  We find that
for metallicities below
$\sim 10^{-2} Z_{\sun}$, where radiation forces onto dust grains in the
flow are negligible, a slightly metal-rich gas favors 
continued accretion and the formation of very massive stars.

\end{abstract}

\keywords{cosmology: theory --- early universe --- galaxies: formation ---
stars: Population III  --- stars: formation}

\section{Introduction}
The problem of the characteristic mass of the first stars that formed in the
universe is still largely unsolved.  Recent theoretical and numerical studies
are revealing in leaps and bounds the physical state of star forming
regions within the first cosmic structures.  In the framework of cold dark
matter (CDM) cosmology smaller objects tend to collapse earlier.  In these
models, a moderately rare object of $\sim 3\sigma$ overdensity collapses and
virializes at about $z \sim 30$.  However, only sufficiently massive objects
can cool in a Hubble time and become luminous by forming stars.  The minimum
cooling mass at $z \sim 30$ is $M_{\rm min}\sim 10^{6}M_{\sun}$, and is a
decreasing function of the collapse redshift (e.g., Haiman, Thoul, \& Loeb
1996; Tegmark et al. 1997; Fuller \& Couchman 2000).  After virialization,
objects more massive than $M_{\rm min}$, called ``first objects'', can
collapse gravitationally owing to molecular hydrogen cooling (e.g., Palla,
Salpeter, \& Stahler 1983, hereafter PSS), and fragment into high-density
clumps.  According to numerical studies, the mass scale of these clumps is
rather high, of the order of $10^3 M_{\sun}$ (Abel, Bryan, \& Norman 2000,
2002; Bromm, Coppi, \& Larson 1999, 2002; Tsuribe \& Inutsuka 2001). However,
the formation of fragments of lower mass, down to $\la 1 M_{\sun}$, is also
possible (e.g., Yoshii \& Saio 1986; Uehara et al. 1996; Nakamura \& Umemura
1999, 2001, 2002; Uehara \& Inutsuka 2000; Omukai 2001).  Indeed, the recent
discovery of HE0107-5240, a star of 0.8 $M_{\sun}$ in the galactic halo 
with virtually no metals ([Fe/H]=$-$5.3), attests the ability of very 
metal-poor gas clouds to give birth to low-mass stars (Christlieb et al. 2002).

After fragmentation, the evolution of gravitationally unstable clumps 
proceeds in a highly
non-homologous fashion, with the central parts collapsing first.  This
runaway phase is induced by cooling provided by H$_2$ line radiation at 
densities $n\la10^{14}\, {\rm cm^{-3}}$, and by H$_2$ collision-induced
emission at higher densities.  The resulting gas temperature is nearly
constant at several $10^2$~K, and the innermost region of $\sim 1M_{\sun}$
becomes fully molecular due to the three-body reaction (PSS).  At densities
$n\sim 10^{16}\, {\rm cm^{-3}}$, the cloud becomes optically thick to
collision-induced absorption and H$_2$ dissociation works as an effective
cooling agent.  Finally, without further cooling mechanism, at $n \simeq
10^{22} {\rm cm}^{-3}$, a small hydrostatic core of mass $\sim
10^{-3}M_{\sun}$ is formed (Omukai \& Nishi 1998), similar to the conditions
found in studies of the collapse of clouds of standard solar composition,
i.e., the stellar (or second) core according to Larson (1969).

In the primordial clouds described by the numerical simulations, the outcome
of the runaway collapse is a tiny ($\sim 10^{-3} \,M_{\sun}$) protostar,
surrounded by a large amount of reservoir gas ($\sim 10^{3} \,M_{\sun}$).  As
a result, the protostar can grow by several orders of magnitude in mass by
accreting the envelope matter.  The mass accretion
rate onto the protostar is determined by the radial density distribution at
the time of core formation, and is related to the prestellar temperature or,
equivalently, to the effective sound speed $c_{\rm s}$ by (e.g., Stahler,
Shu, \& Taam 1980):

\begin{equation} 
\dot{M}_{\rm acc}\,\sim\, \frac{c_{\rm s}^{3}}{G}.  
\label{eq:mdot} 
\end{equation}

The evolution of zero-metal stars in the main accretion phase was studied by
Stahler, Palla, \& Salpeter (1986; hereafter SPS) up to 10.5~M$_{\sun}$ and
by Omukai \& Palla (2001; hereafter, Paper I) for higher masses under a
constant mass accretion rate of $4.41 \times 10^{-3} M_{\sun}$~yr$^{-1}$,
corresponding to a gas temperature of 1700~K in eq.~(1).

The mass of the forming star is set at the time when protostellar accretion
stops.  In paper I, we have found that fast accreting protostars enter a
phase of rapid expansion at a mass of $\sim 300~M_{\sun}$ when the
luminosity becomes very close to the Eddington limit. This event may
determine the onset of a powerful stellar wind driven by radiation pressure
that can effectively quench further accretion of circumstellar material.  In
that event, after a transient internal readjustement, the protostar is
expected to settle as an ordinary zero-age main sequence (ZAMS) star.

The upper limit of $\sim 300~M_{\sun}$ to the protostellar mass has been
obtained using a constant value of the mass accretion rate.  In realistic
collapse calculations, deviations from the isothermal assumption are expected
to occur, and even in the ideal case the mass accretion rate is found to vary
in time (e.g., Foster \& Chevalier 1993; Tomisaka 1996).  Therefore, it is
important to extend our initial study to a larger set of conditions in order
to explore the robustness of the results described above.  Along with 
differences in the mass accretion rate, we will also address the issue
of the impact of a small abundance of heavy elements.  It is expected
that metals may alter the mass limit drastically as a result of the existence
of a dust shell and because of steller evolution effects.

In \S 2, we sketch our numerical approch. Then, in \S 3 we describe the
results of the main accretion phase obtained with a constant and
time variable $\dot M_{\rm acc}$. The effects of a finite, non-zero
metallicity on protostellar evolution are discussed in \S 4.  In \S
5 we discuss the implications of our work.
Finally, in \S 6, we give the conclusions.

\section{Numerical Approach}
The basic strategy and equations used here are the same as in SPS. In that
scheme, the protostellar evolution is treated as a sequence of steady state
accretion flows onto a growing core (see Figure~1 of SPS for a definition of
the various regions). The core is assumed to be in hydrostatic equilibrium
and the ordinary stellar structure equations are applied.  If the gas in
front of the accretion shock is optically thin, no envelope model is
constructed and the boundary condition given by eq. (4a) of SPS is applied at
the core surface. In case the preshock gas is optically thick, we integrate
the equations inside the radiative precursor from the photosphere to the core
surface (see eqs. 7a-d of SPS).  Outside the photosphere, we assume a
free-falling flow.  The main difference with the SPS work is in the use of
updated opacity tables: for $T<7000$~K, we take the results of Lenzuni,
Chernoff, \& Salpeter (1991) for a gas composition $X=0.72, Y=0.28$, and the
OPAL opacity at higher temperatures (Iglesias \& Rogers 1996), with a
slightly different hydrogen abundance $X=0.70$).  For partially polluted gas,
we have used the results of Alexander \& Ferguson (1994) with $X=0.70, Y=0.30$
for $T<7000$~K, and the OPAL opacity for higher temperatures.  Although the
primordial He abundance is about $Y=0.24$ (e.g., Galli et al. 1995; 
Izotov \& Thuan 1998), we used rather high value of $Y \ga 0.28$ as stated 
above because of the lack of opacity tables for primordial gas.  This
causes no significant error.

The calculations are started at the protostellar mass of $0.01M_{\sun}$.
The initial models are arbitrary and constructed using an entropy 
distribution of the core given by   
\begin{equation}
s(M)=s_{\rm 0}+\beta (M/0.01M_{\sun})^{2}\,,
\end{equation}
where $M$ is the enclosed mass within a spherical surface, 
$s_{\rm 0}=-13.05$ and
$\beta=7.488$ in units of $\cal R$, and the zero-point is arbitrarily set at
$T=2.05 \times 10^{5}$~K and $\rho=5.16$~g~cm$^{-3}$.  With the mass
accretion rate used by SPS, $\dot{M}_{\rm acc}=4.41 \times
10^{-3}$~M$_\sun$~yr$^{-1}$, the initial central temperature is $10^{5}$~K.
This value increase (decreases) for higher (lower) accretion rates.  The
initial conditions do not affect the later evolution, since the protostar
adjusts quickly during the ``decay-of-transient phase'' (for the protostellar
mass $M_{\ast}< 0.1M_{\sun}$; SPS) to the state appropriate to an accreting
object.

\section{Formation of Primordial Stars:\\ 
Effects of Variation of the  Mass Accretion Rate}

\subsection{Evolution under different mass accretion rates}
First, we study the effects of using different, but constant values of
$\dot{M}_{\rm acc}$ on the evolution of primordial protostars.  We define
$\dot{M}_{\rm fid}=4.41 \times 10^{-3}~M_\sun$~yr$^{-1}$ as the fiducial
value of the accretion rate used by SPS and in Paper I.  To follow the
dependence of the evolution on $\dot{M}_{\rm acc}$, we have studied four
cases, varying $\dot{M}_{\rm acc}$ from $1/4\, \dot{M}_{\rm fid}$ to $2\,
\dot{M}_{\rm fid}$, thus covering the range $1.1-8.8 \times
10^{-3}~M_\sun$~yr$^{-1}$.

The evolution of the protostellar radius $R_{\ast}$ as a function of 
the accretion rate is shown in Figure~\ref{fig:MR_prm}.  As the calculations
start from rather arbitrary initial conditions, the protostar adjusts itself
to the state appropriate to an accreting object by entropy redistribution.
As a result, the fast entropy variations cause the bumps in radius seen in
Figure~\ref{fig:MR_prm} at masses $\leq$0.03 $M_\sun$.  After this minor
transient phase, the evolution proceeds smoothly and the similarity of the
curves allows the identification of three distinct phases. The first two are
characterized by a common trend where the protostellar radius $R_{\ast}$
increases almost linearly with the protostellar mass $M_{\ast}$ in the
logarithmic scale, reaches a peak between $\sim$10 and 20 $M_\sun$, and then
drops substantially.  The adiabatic accretion phase (Phase I, $M_{\ast} \la$
a few to $\sim 10 M_{\sun}$) is marked by a gradual expansion of the radius
and lasts until the arrival of the internal luminosity wave at the surface,
whose eruption causes the sudden swell of the external layers (e.g., Palla \&
Stahler 1991).  In the Kelvin-Helmholtz contraction phase (Phase II, a
few-$10 M_{\sun} \la M_{\ast} \la 30-60 M_{\sun}$) the increased
gravitational pull of the growing star stops the expansion and forces the
star to contract toward the conditions appropriate for a main sequence
object.  The third phase varies markedly, depending critically on the adopted
value of $\dot{M}_{\rm acc}$.

If the rate is comparable to $\dot{M}_{\rm fid}$ radiation pressure causes
the abrupt expansion of the surface layers (Phase IIIb) seen in the two upper
curves of Figure~\ref{fig:MR_prm} computed for $\dot{M}_{\rm acc}=(1-2)\,
\dot{M}_{\rm fid}$.  The protostellar radius increases tremendously and the
impulsive release of energy likely causes the stripping of the external
layers of the protostar by a radiation-driven stellar wind. As shown in
Paper~I, this event signals the end of the accretion phase and sets the
maximum value of the protostellar mass.  In the opposite case where
$\dot{M}_{\rm acc} < \dot{M}_{\rm fid}$, contraction proceeds unimpeded by
accretion.  In the cases shown in Figure~\ref{fig:MR_prm} and computed for
$\dot{M}_{\rm acc}=1/2$ and $1/4 \dot{M}_{\rm fid}$, the protostar reaches
the ZAMS at a mass of $\sim 50~M_\sun$ (Phase IIIa).

The global evolution of the core interior is shown in
Figure~\ref{fig:core_qrt} for the low mass accretion rate ($\dot{M}_{\rm
acc}=1/4 \dot{M}_{\rm fid}$) and in Figure~\ref{fig:core_fid} for the case
$\dot{M}_{\rm acc}=\dot{M}_{\rm fid}$.  In these figures, the bottom panels
illustrate the evolution of the central variables, as well as those of the
position of the temperature maximum for the same cases.

The overall evolution of the interior luminosity $L_{\ast}$ is shown 
in the top panel
for $\dot{M}_{\rm acc}=1/4 \dot{M}_{\rm fid}$ and in the bottom panel of
Figure~\ref{fig:Lnuc} for $\dot{M}_{\rm acc}=2 \dot{M}_{\rm fid}$.  The
contribution to the total luminosity provided by nuclear reactions of
deuterium and hydrogen (through the $pp$-chain and CN-cycle) at the center of
the protostar is also indicated. Notice how the onset of each burnig stage is
shifted to higher masses for bigger values of $\dot{M}_{\rm acc}$.

Having sketched the main features of the evolution, we now describe in
detail the results of each evolutionary phase identified above.

\subsubsection{Phase I: adiabatic accretion and propagation of the luminosity 
wave}

Early in the evolution, the opacity in the interior is due to free-free
absorption ($\kappa_{\rm ff} \propto \rho T^{-3.5}$) and is very high
due to the low temperatures ($T<10^{6}$~K). As shown in Figure~\ref{fig:Lnuc},
the luminosity is extremely low and consequently the cooling time of the
internal regions (i.e., the Kelvin-Helmholtz time $t_{\rm KH} =
GM_{\ast}^{2}/R_{\ast}L_{\ast}$) is longer than the evolutionary timescale
(i.e., the accretion time $t_{\rm acc}=M_{\ast}/\dot{M}_{\rm acc}$).  As a
consequence, the accreted material piles up without further cooling, after
passing through the accretion shock and the settling region.  During this
adiabatic accretion phase, the protostar swells gradually as the mass
increases.  The protostellar radius obeys the relation (see equation 21 of SPS)
\begin{equation} R_{\ast} \propto M_{\ast}^{0.27}\, \dot{M}_{\rm acc}^{0.41},
\end{equation} that can be derived assuming that the luminosity is dominated
by the accretion contribution and that the opacity in the envelope is due to
H$^{-}$ bound-free absorption, scaling as $\kappa \propto T^{14.5}$.

The increase of the temperature accompanied by the growth of the 
protostellar mass produces a significant decrease of the opacity.
Eventually, the heat contained inside 
the protostar begins to propagate outward as a luminosity wave 
(see Figure~10 of SPS). Then, the interior luminosity $L_{\ast}$ 
increases dramatically 
at a mass of $\sim4 M_{\sun}$ for $\dot{M}_{\rm acc}=1/4 \dot{M}_{\rm fid}$, 
and at $\sim8M_{\sun}$ for $\dot{M}_{\rm acc}= \dot{M}_{\rm fid}$ 
(see Figure~\ref{fig:Lnuc}).
The arrival of the luminosity wave at the surface produces
the sudden swelling of the protostar, marking
the end of Phase I.
In the course of this expansion, a short-lived surface convective
region develops which extends to about 20\% in mass (Figs.~\ref{fig:core_qrt}, 
\ref{fig:core_fid}).

In the case of low $\dot{M}_{\rm acc}$ (Figure~\ref{fig:core_qrt}), 
we can see a
temporary, but active phase of deuterium burning around $3M_{\sun}$, when the
nuclear energy generation rate exceeds $0.1L_{\ast}/M_{\ast}$. This episode
is just the consequence of the low value of $L_{\ast}$ before the arrival of
the luminosity  wave (see the top panel of Figure~\ref{fig:Lnuc}).  
However, the deuterium
luminosity $L_{\rm D}$ is very small and deuterium is hardly consumed with no
significant consequences on the evolution.

As a result of the outward increase of the entropy, the temperature is
highest off-center, at an internal  mass of $M/M_{\ast}\sim 0.1$.  The
position of the off-center temperature peak is shown in the top panels of
Figs.~\ref{fig:core_qrt} and \ref{fig:core_fid} and is maintained throughout
the adiabatic accretion phase.

\subsubsection{Phase II: Kelvin-Helmholtz contraction}

After the relaxation of the interior and the passage of the luminosity wave, 
the major opacity source in protostars is provided by electron scattering, 
except in a small central region 
where free-free absorption is dominant, and in a thin surface layer 
characterized by bound-free absorption of H and He.

The condition of radiative equilibrium and of an opacity independent of
density and temperature throughout most of the interior determines the 
well known mass--luminosity
relation $L_{\ast} \propto M_{\ast}^{3}$,
and the Kelvin-Helmholtz time becomes
\begin{equation}
t_{\rm KH} = \frac{GM_{\ast}^{2}}{R_{\ast}L_{\ast}} 
           \propto \frac{1}{M_{\ast}R_{\ast}}\,.
\end{equation}
If the radius is sufficiently large, $t_{\rm KH}<t_{\rm acc}$ and the core
shrinks due to fast energy loss, while in the opposite case the core swells
adiabatically.  The radius of the protostar actually adjusts towards equality
of the two timescales $t_{\rm KH} \simeq t_{\rm acc}$, leading to the
relation 
\begin{equation} 
R_{\ast} \propto \dot{M}_{\rm acc}/M_{\ast}^{2}.
\end{equation} 
Small deviations from this relation are caused by minor
contributions to the opacity from processes other than electron scattering.

Figure~\ref{fig:ML} shows the evolution of the interior luminosity to mass
ratio, $L_{\ast}/M_{\ast}$, as a function of the protostellar mass
$M_{\ast}$. During KH contraction, $L_{\ast}/M_{\ast}$ increases rapidly and
the mass-luminosity relation follows approximately a unique relation
independent of the value of the mass accretion rate, as expected in the ideal
case of constant opacity. Also, the photospheric
luminosity, which is the sum of the interior ($L_{\ast}$) and accretion
($L_{\rm acc}$) luminosity, is independent of $\dot{M}_{\rm acc}$, since
$L_{\rm acc} \propto \dot{M}_{\rm acc}/R_\ast$ and $R_\ast \propto
\dot{M}_{\rm acc}$.

At the beginning of contraction, the maximum temperature exceeds $10^{6}$K,
and deuterium can start burning.  Because of the off-center temperature peak,
burning propagates both inward and outward, scouring out all of the available
deuterium (see Fig.s \ref{fig:core_qrt} and \ref{fig:core_fid}).  
Under the high accretion rates considered here, however,
D-burning plays a minor role in the evolution and contributes at most
$\sim$1/4 of the interior luminosity (see Figure \ref{fig:Lnuc}).  
The off-center temperature peak is only a transient
property of the core and disappears at $20M_{\sun}$ for $\dot{M}_{\rm acc}=1/4
\dot{M}_{\rm fid}$, and $40M_{\sun}$ for 
$\dot{M}_{\rm acc}= \dot{M}_{\rm fid}$, as
seen in Figs.~\ref{fig:core_qrt} and \ref{fig:core_fid}.

After deuterium is completely destroyed and the protostar has contracted
further, the $pp$-chain becomes operative in the center.  However, even in
this case, the energy generated by these reactions is not enough to
contrast contraction effectively, owing to the saturation of the energy
generation rate at high temperatures (see Figure~\ref{fig:Lnuc}).  
Therefore, despite nuclear heating, protostars can continue
the phase of gravitational contraction.

\subsubsection{Phase IIIa: Settling onto the ZAMS with continuing accretion 
($\dot{M}_{\rm acc} \leq 1/2 \dot{M}_{\rm fid}$) }

For accretion rates of $\dot{M}_{\rm acc}=1/2$ and 1/4$\dot{M}_{\rm fid}$,
during contraction the central temperature becomes high enough to synthesize
carbon via the CN-cycle and H-burning becomes a significant source of
luminosity.  Figure~\ref{fig:ML} shows that at
this time gravitational contraction is halted and the interior luminosity to
mass ratio $L_{\ast}/M_{\ast}$ stops increasing.  The protostar relaxes
quickly to the ZAMS and the models converge toward a single
mass-luminosity-radius relation, as shown in Fig.s~\ref{fig:MR_prm} and 
\ref{fig:ML}.  With
H-burning working as a thermostat, the central temperature remains almost
constant at about $10^{8}$ K 
(see the bottom panel of Figure~\ref{fig:core_qrt}).

At this point, the interior luminosity is essentially supplied by nuclear
reactions (see the top panel of Figure \ref{fig:Lnuc}).  
Since part of this energy source is
used to heat the interior, the specific entropy in the convective core
increases as the stellar mass grows.  This results in the decrease of the
central density, while the temperature remains constant.  The mass fraction
of the convective core increases (see the top panel of 
Figure~\ref{fig:core_qrt}) and the CN
abundance reaches very quickly a level of $\sim 10^{-9}$, thereby gradually
increasing (see the bottom panel of Figure~\ref{fig:core_qrt}).

As shown in Figure~\ref{fig:ML},  throughout the evolution the luminosity
remains below the Eddington value by a rather large margin.  The latter has
been computed using electron scattering as the main source of opacity in the 
accretion flow.  As a
result, accretion can continue without experiencing the effects of the
retardation force exerted onto the ionized radiative precursor. Therefore,
the upper limit of the protostellar mass formed in these conditions is set by
the amount of circumstellar material left during accretion and not by the
feedback effect of radiation pressure.

\subsubsection{Phase IIIb: Rapid expansion due to radiation pressure 
($\dot{M}_{\rm acc} \geq \dot{M}_{\rm fid}$) }
 
For high accretion rates, the protostar undergoes a phase of violent
expansion as soon as the interior luminosity becomes comparable to the
Eddington limit. This process is caused by the strong radiation force exerted
both on the stellar surface and the radiative precursor.  The result is that
fast accreting protostars never converge to an ordinary ZAMS structure.
Let us see more in detail how this condition is achieved.

During the contraction phase, the luminosity-to-mass ratio increases until
the conditions at the center are appropriate to start nuclear burning (see
Fig. \ref{fig:ML}). This happens at progressively large masses as the
accretion rate increases.  For sufficiently high $\dot{M}_{\rm acc}$, the
total luminosity, which is the sum of the interior and accretion luminosity, 
reaches the Eddington limit before the ignition of hydrogen.
This is the case for $\dot{M}_{\rm acc}=2 \dot{M}_{\rm fid}$, while for the 
lower value of $\dot{M}_{\rm acc}=\dot{M}_{\rm fid}$ the situation is more 
complicated. In the accreting envelope, gravitational acceleration is 
approximately balanced by the deceleration due to the radiation force.  
Then, the ram pressure onto the protostellar surface decreases suddenly, and 
the protostar begins to expand without violating the Eddington limit by 
lowering the accretion luminosity $L_{\rm acc} \propto R_{\ast}^{-1}$. During
expansion, the temperature drops, while the radiation force and the opacity
increase at the surface, the latter peaking sharply owing to the contribution
of ionization of trace amount of atoms.  This is mostly a surface effect
since the main opacity source is still electron scattering both inside the
protostar and in the accreting envelope.  By this mechanism, the expansion is
accelerated and possibly results in the stripping of the surface layers and
in the reversal of the material in the accreting envelope.  Thus, the main
phase of accretion is over and the protostar can rapidly relax onto a ZAMS
structure. According to the results shown in Fig.~\ref{fig:MR_prm}, 
this critical phase is achieved at a mass of $\sim 90 M_\sun$ for
$\dot M_{\rm acc}=2 \dot{M}_{\rm fid}$, and $\sim 300 M_\sun$ for
$\dot M_{\rm acc}= \dot{M}_{\rm fid}$. Therefore, {\it our models show that
the upper limit of the mass shifts toward smaller values for progressively
higher values of the mass accretion rate.}

\subsubsection{A general consideration}
 
We have seen that, depending on the actual value of the accretion rate,
the final fate of primordial protostars can be quite different and
leads to strongly different values of the maximum limit of their mass.
This suggests to look deeper into the physical origin of the 
bifurcation displayed in the mass-radius relation of Figure~\ref{fig:MR_prm}.

The total luminosity of a protostar, $L_{\rm tot}$, consists of the interior, 
$L_{\ast}$, and accretion luminosity, $L_{\rm acc}$:
\begin{equation}
L_{\rm tot}=L_{\ast}+L_{\rm acc} \simeq L_{\ast}+GM_{\ast}\dot{M}_{\rm acc}/R_{\ast}. 
\end{equation} 
Imagine that the protostar reaches the ZAMS star while continuing accretion,
as in the cases of $\dot{M}_{\rm acc} \leq 1/2 \dot{M}_{\rm fid}$.
Since a unique mass-luminosity-radius relation
holds for ZAMS stars, the total luminosity would be 
\begin{equation}
L_{\rm tot} \simeq L_{\rm ZAMS}+GM_{\ast}\dot{M}_{\rm acc}/R_{\rm ZAMS}\,,
\label{eq:Ltot}
\end{equation} 
where we have set 
$R_{\ast}=R_{\rm ZAMS}(M_{\ast})$ and $L_{\ast}=L_{\rm ZAMS}(M_{\ast})$.
If the accretion rate is larger than the critical value
\begin{equation}
\dot{M}_{\rm crit}=\frac{4 \pi c R_{\rm ZAMS}}{\kappa_{\rm es}}\,
\left(1-\frac{L_{\rm ZAMS}}{L_{\rm Edd}}\right)\,
 \simeq 4 \times 10^{-3} M_{\sun}/{\rm yr}\,,
\label{eq:Mdot_cr}
\end{equation}
where $\kappa_{\rm es}$ is the electron scattering opacity,
the total luminosity exceeds the Eddington limit $L_{\rm Edd}$, implying that
the star cannot adjust to the ZAMS structure at these accretion rates.
Although the critical accretion rate is a function of the mass of 
the central star, the dependence is weak considering the 
$M_{\ast}-L_{\ast}-R_{\ast}$ relations.

Note that the critical accretion rate $\dot{M}_{\rm crit}$ is the 
value where the total luminosity $L_{\rm tot}$ reaches $L_{\rm Edd}$ 
at the same time as the protostar reaches the ZAMS.
Thus, we can derive $\dot{M}_{\rm crit}$ by using 
the ZAMS quantities, although protostars start expanding 
before reaching the ZAMS for $\dot{M}_{\rm acc} > \dot{M}_{\rm crit}$.
For $\dot{M}_{\rm acc}<\dot{M}_{\rm crit}$, $L_{\rm tot}$ remains
below $L_{\rm Edd}$ up to the point where the protostar reaches the ZAMS, 
while for $\dot{M}_{\rm acc}>\dot{M}_{\rm crit}$, $L_{\rm tot}$ equals
$L_{\rm Edd}$ in the course of the Kelvin-Helmholtz contraction and
the protostar starts expanding.

The weak dependence of $\dot{M}_{\rm crit}$ on the protostellar mass $M_{\ast}$
can be seen more clearly if we define the Eddington radius 
as the radius of the core where the total luminosity (defined in
eq. \ref{eq:Ltot}) equals the Eddington limit $L_{\rm Edd}$:
\begin{equation}
R_{\rm Edd} = \frac{\kappa_{\rm es} \dot{M}_{\rm acc}}{4 \pi c (1-\Gamma)}\,, 
\end{equation}
where $\Gamma\equiv L_{\ast}/L_{\rm Edd}$.

In the two panels of Figure~\ref{fig:MRedd}, 
we show the Eddington radius and the core radius 
for the four values of the accretion rate.
In all cases, when the core mass is small and $\Gamma \ll 1$, 
the Eddington radius is constant. Later on, the evolution differs.
Consider first the case for low $\dot{M}_{\rm acc}$.
During the contraction phase, the luminosity-to-mass ratio increases, the
core shrinks, but the Eddington radius starts a slow rise.
However, $R_{\rm Edd}$ never climbs up to $R_\ast$, since the core meets
the ZAMS conditions before and its radius also increases as a result
of the heating provided by nuclear burning. Thereafter, $R_{\rm Edd}$
remains below the ZAMS radius and accretion can proceed unimpeded
by radiation forces. Since the $R_{\rm Edd}/R_\ast$ ratio remains
approximately constant, the critical accretion rate is almost 
independent of the protostellar mass.

If, on the other hand, the protostar reached the ZAMS structure with an
accretion flow $\dot{M}_{\rm acc}> \dot{M}_{\rm crit}$, the total luminosity
would exceed the Eddington limit.  Unlike the previous case, at some point of
the evolution the Eddington and core radii would become equal (see the bottom
panel of Fig. \ref{fig:MRedd}).  The core now undergoes the runaway
expansion, keeping the luminosity slightly below the Eddington limit. The
increased radius reduces the accretion luminosity, thus allowing more gas to
flow for a short period.

\subsubsection{Properties of the radiative precursor}

Except for the short-lived phase of the peak radius around $10 M_{\sun}$, 
primordial protostars are embedded within an optically thick envelope.  The
spatial extent of the precursor, which is the optically thick part of the 
accretion flow, corresponds approximately to that of the
ionized region.  However, the ionized region around the accreting 
protostar differs from an ordinary HII region in that it is opaque to optical 
photons, and both matter and radiation are in the thermodynamic equilibrium.

In Figure~\ref{fig:MRph}, we show the evolution of the photospheric radius
for two cases with  $\dot{M}_{\rm acc}=1/4 \dot{M}_{\rm fid}$ and
$\dot{M}_{\rm fid}$. During the adiabatic phase, most of the luminosity is
generated at the accretion shock and the ratio of the photospheric to core
radius remains nearly constant, with a value of $\sim$1.4 (see also eq. 19 of
SPS).  After the luminosity wave reaches the surface, the interior luminosity
exceeds the accretion contribution. As a result of core contraction and
of the increased heat deposit in the precursor, the
ratio of photospehric to core radius increases dramatically. The trend
is reversed during the final stages of the evolution.

The evolutionary tracks of the photosphere in the HR diagram
are displayed in Figure~\ref{fig:HR} for the four accretion histories.  For
comparison, we also show the evolution of the protostellar surface 
and locus of the ZAMS for metal-free stars.
The distinction between the surface temperature and the ZAMS effective 
temperature after reaching the MS phase is a consequence of the difference in 
boundary conditions at the stellar surface. 
Because of the optically thick envelope, the hot surface of the protostar
remains invisible in the optical.  The photospheric temperature is locked at
$\sim$6000 K due to strong sensitivity of the H$^{-}$ bound-free opacity on
temperature.  However, for low $\dot M_{\rm acc}$, the tracks make an
abrupt shift toward higher temperatures for $M_{\ast} \ga 100 M_{\sun}$.  As
the protostellar mass increases, the density in the accreting envelope
decreases, while the temperature and ionization degree both increase.
Electron scattering replaces the H$^{-}$ b-f absorption as the major source
of opacity at the photosphere.  Since this process is independent of
temperature, the effective temperature rises as the density in the accreting
envelope decreases.  Correspondingly, the photospheric radius decreases
sharply (Figure~\ref{fig:MRph}). The tracks would then join the corresponding
ZAMS models at masses exceeding $1000 M_\sun$, but we cannot determine
exactly where this would occur since such models are not available in the 
literature.

\subsection{Protostellar Evolution with a mass dependent accretion rate}

The results described so far have been obtained under the rather limiting
assumptions that the accretion rate is constant throughout the evolution and
of large magnitude ($\ga 10^{-3} M_{\sun} {\rm yr}^{-1}$), 
at least two orders of magnitude more
than what is normally considered in the collpase phase of present-day
molecular cores.  That $\dot M_{\rm acc}$ is so large depends on the thermal
and chemical properties of the primordial gas, which is known to be a very
poor radiator owing to the lack of molecules and dust grains. Since the gas
temperature during collapse is mainly determined by the excitation of H$_2$
molecules, it is natural to expect that collapse models find
typical values well in excess of several hundred degree, thus justifying the
large values of $\dot M_{\rm acc}$ through eq.~(1).  On the other hand, the
use of a single value of the accretion rate is less motivated since, as we
have noted in \S 1, numerical simulations have shown a rather
strong dependence on time of the amount of material that can actually
collapse, even in the case of strict isothermality.

Thus, it is natural to ask how robust is the sequence of events that we
have described in the previous sections in view of a more general behavior of
the gas dynamics, and hence of the time history of the accretion rate.  Since
the key factor that determines the ultimate fate of massive protostars is
whether $\dot M_{\rm acc}$ is above or below the critical value $\dot{M}_{\rm
crit} \simeq 4 \times 10^{-3}~M_\sun$~yr$^{-1}$, we should consider how and
when this situation is met under more realistic conditions.

A useful guide is provided by the recent 3D hydrodynamical simulations of the
formation of the first stars presented by Abel, Bryan \& Norman (2002;
hereafter ABN).  Starting from cosmological initial conditions, these authors
have followed the evolution of a collapsing primordial clump up to the stage
when optical depth effects become very large, i.e., when the central number
density reaches $\sim 10^{12} {\rm cm^{-3}}$.  Due to numerical limitations
and their optically thin approximation for H$_2$ line radiation, ABN could
not follow correctly the subsequent history of protostar formation and
accretion phases. However, they provide the run of the accretion time as a
function of enclosed gas mass which can be converted into an effective
accretion rate that depends on the growing protostellar mass.  Using the
results shown in Fig. 5 of ABN, we obtain the following fit
\begin{equation}
\dot{M}_{\rm ABN}= \left\{ 
\begin{array}{lc}
7.76 \times 10^{-3}\, (M_{\ast}/M_{\sun})^{-0.24}\,  M_{\sun}\,{\rm yr}^{-1} 
& 0.6\,M_\sun< M_{\ast}<13 \,M_{\sun} \\
1.51 \times 10^{-3}\, (M_{\ast}/M_{\sun})^{0.41}\,   M_{\sun}\,{\rm yr}^{-1}
& 13\,M_{\sun}<M_{\ast}<60\,M_{\sun} \\
11.2\, (M_{\ast}/M_{\sun})^{-1.76}\,   M_{\sun}\,{\rm yr}^{-1}
& 60\,M_{\sun}<M_{\ast}. 
\end{array}
 \right.
\label{eq:abn}
\end{equation}
In order to avoid computational difficulties,
we have used $\dot{M}_{\rm acc}=8.8 \times 10^{-3}
M_{\sun}$~yr$^{-1}$ for lowest mass models ($M_{\ast}<0.6~M_{\sun}$). 
Thus, the accretion rate starts at very high values, it remains well above
$\sim$10$^{-3}~M_\sun$~yr$^{-1}$ up to about 60 $M_\sun$, 
and then falls rapidly for very massive stars. Most importantly,
$\dot{M}_{\rm ABN}$ drops below the critical value for 
$M_{\ast}>95~M_{\sun}$.

The evolution of the protostar with the accretion rate given by
equation~(\ref{eq:abn}) is displayed in Figure~\ref{fig:MR_abn}. 
For comparison, we
also show the results at fixed $\dot{M}_{\rm acc}$ discussed before.  In
Figure~\ref{fig:MR_abn2}, the evolution of the photospheric radius is
compared to that of the core.  During the initial adiabatic phase, the
protostar expands more slowly than in the constant $\dot{M}_{\rm acc}$ case
in response to the decrease of $\dot{M}_{\rm ABN}$ for
$M_{\ast}< 13~M_{\sun}$.  The thermal relaxation and the appearance of the
luminosity wave with the expansion of the radius occur at approximately the
same mass as for $\dot{M}_{\rm acc}=\dot{M}_{\rm fid}$, since the
instantaneous value of $\dot{M}_{\rm ABN}$ is nearly the same as the fiducial
one ($0.97$~$\dot M_{\rm fid}$ at 13 $M_{\sun}$).

During Phase II, the protostar contracts more gradually than in the fiducial
case since $\dot{M}_{\rm ABN}$ reaches a minimum at $13~M_\sun$
and increases thereafter. The core shrinks down to its minimum value of
$\sim 20~R_\sun$ at about 50 $M_\sun$ and then rebounds when radiation
pressure effects begin to appear. This is the most delicate part of the
evolution that depends critically on the behavior of $\dot{M}_{\rm ABN}$.
As Fig.~\ref{fig:MR_abn} shows, the phase of violent expansion is avoided
in this case and the protostar resumes further contraction.
At about $90~M_{\sun}$, hydrogen is ignited at the center and 
a convective core begins to develop. Then, 
the accretion rate falls below the critical value at $95~M_{\sun}$.
Shortly after, the protostar relaxes to a ZAMS star, without suffering
from surface expansion or a blow out of the accretion envelope.
Therefore, accretion is expected to continue unimpeded to very high masses.

With continuing accretion, 
the final mass of the forming star is set by how much material can be accreted
during the stellar lifetime (e.g., Larson \& Starrfield 1971).
Because of the sharp decrease of the accretion rate, 
the accretion time $t_{\rm acc}=M_{\ast}/\dot{M}_{\rm ABN} \simeq 
3 \times 10^{4} (M_{\ast}/100M_{\sun})^{2.76} {\rm yr}$ 
(for $M_{\ast}>60M_{\sun}$)
exceeds the stellar lifetime ($\sim 3 \times 10^{6}$ yr) 
at $M_{\ast} \simeq 500M_{\sun}$.
Therefore, even if accretion goes on throughout the entire stellar lifetime, 
the stellar mass would not exceed $500M_{\sun}$ by a large margin.
A more detailed analysis has concluded that if the accretion continues, 
the mass attained during the entire stellar lifetime is about 
$600M_{\sun}$ (ABN).
In reality, when the accretion rate drops below a certain value, 
the star is hot enough to produce a copious amount of ionizing photons that
create an ordinary HII region, whose
expansion can effectively reverse the infalling gas.
Omukai \& Inutsuka (2002) have shown that the mass limit set by this
process is very similar to the limit imposed by the main-sequence lifetime
($\la 460M_{\sun}$; see their eq.~27). 

Returning to the evolution of the photospheric radius, in the high mass
regime ($M_{\ast}>60M_{\sun}$), the rapidly decreasing accretion rate
causes the abrupt turn around shown in Fig.~\ref{fig:MR_abn2}, after reaching
the peak value of $\simeq 1000~R_{\sun}$ at about $100~M_{\sun}$.  At a mass
of $\sim 500~M_{\sun}$, the radiative precursor disappears and the accretion
envelope becomes optically thin:  At this point the hot stellar surface
becomes directly visible from outside.

\section{The formation of second-generation stars:\\ Effects of metal 
enrichment}
Second-generation stars are formed out of material that has been contaminated
by metals released from Population III SNe. Heavy elements affect the
chemistry and dynamics of the gas in several ways.  For example, detailed
models have shown that a metal abundance of $\sim 10^{-2}~Z_{\sun}$ can
substantially change the cooling properties of gas, thereby lowering the
minimum mass of cosmological objects (e.g., Fall \& Rees 1985).  A metal
abundance as low as $\sim 10^{-4}~Z_{\sun}$ can influence the
fragmentation mass scale of star-forming clouds (Omukai 2000; Bromm et al.
2001a; Schneider et al. 2002).  Protostellar evolution is also altered
through important changes in the opacity and in the initial abundance of the
CN elements.  Therefore, it is important to consider how even a minute
enrichment of heavy metals modifies the picture of star formation that we
have been describing so far.

Since the preceding discussion has demonstrated that the qualitative evolution
of accreting protostars is the same for fixed and variable mass accretion rates,
we have run several models at constant $\dot{M}_{\rm acc}$ for nonzero
metallicities. Figure~\ref{fig:MR_half} shows the resulting mass-radius 
relation obtained with  
$\dot{M}_{\rm acc}=1/2 \dot{M}_{\rm fid}=2.2 \times 
10^{-3}~M_{\sun}$~yr$^{-1}$ for $Z=10^{-6}=5 \times 10^{-5} Z_{\sun}$ 
and $10^{-4}=5 \times 10^{-3} Z_{\sun}$, 
along with the standard zero-metal case.
The evolution is basically the same, except for
the temperature at which the CN-cycle starts operating.
At higher $Z$, significant burning begins at lower temperatures. 
Therefore, the contraction phase is less pronounced, yielding a larger
core radius on the main sequence. This explains the shift of the minimum
radius in Figure~\ref{fig:MR_half} at higher $Z$. 
On the other hand, the protostellar luminosity at a given mass is the same
in all models.  The main effect of metals consists of an increase of the
critical accretion rate that varies from 
$\dot{M}_{\rm crit}=4 \times 10^{-3}$~M$_\sun$~yr$^{-1}$ for $Z=0$ to 
$9 \times 10^{-3}~M_\sun$~yr$^{-1}$ for $Z=10^{-4}=5 \times 10^{-3} Z_{\sun}$. 
The change of a factor $\sim$2 in $\dot{M}_{\rm crit}$ reflects the similar
variation of the protostellar radius (through equation~\ref{eq:Mdot_cr}).  Of
course, a higher value of the critical accretion rate implies that protostars
can keep accreting without entering the phase of violent expansion.

The effect of metals is much more dramatic if we consider other
values of $\dot{M}_{\rm acc}$.
Figure~\ref{fig:MR_fid} displays the mass-radius relation computed with 
$\dot{M}_{\rm acc}=\dot{M}_{\rm fid}=4.4\times 10^{-3}~M_\sun$~yr$^{-1}$
for the three metal compositions.
The accretion rate is now slightly higher than the 
critical value for  the metal-free case, 
but smaller than that at $Z=10^{-6}=5 \times 10^{-5} Z_{\sun}$ 
($6 \times 10^{-3}~M_\sun$~yr$^{-1}$). 
As a result, the violent expansion that occurs at $Z=0$ is no longer found, 
even in the lowest metallicity case.

Thus, we conclude that the presence of metals, by increasing the
critical accretion rate, favors the formation of very massive objects.
This result is not intuitive since metals, locked up in grains, 
are usually believed to reduce the upper mass limit because of
the enhanced radiation force onto the dust shell surrounding 
the protostar (e.g., Kahn 1974; Wolfire \& Cassinelli 1987). 
To understand this behavior, let us remind that the mechanism that should
prevent infall is  the violent expansion of the surface layers 
caused by the radiation force on the ionized 
radiative precursor. In the case of high $Z$,
radiation pressure is imparted onto the dust shell 
which is located at a much greater distance from the precursor.

We can derive a simple estimate of the critical metallicity that 
separates the two regimes.
The radiation force on the dust shell becomes bigger than that 
on the ionized precursor if the opacity in the shell 
$\kappa_{\rm dust}=\kappa_{\rm dust}^{(0)}(Z/Z_{\sun})$, where 
$\kappa_{\rm dust}^{(0)} \simeq$~30~cm$^2$~g$^{-1}$ is 
the present value of dust opacity (e.g., Beech \& Mitalas 1994), 
exceeds that due to electron scattering in the ionized precursor, 
$\kappa_{\rm prec}=$~0.4~cm$^2$~g$^{-1}$.
Here, we have assumed that the dust properties (e.g., composition, 
size distribution, etc.) do not change with metallicity.
In this case, the critical value of metallicity is given by
\begin{equation}
Z>Z_{\rm dr}= 0.01 \,Z_{\sun} \,
\left(\frac{\kappa_{\rm dust}^{(0)}}{30\,{\rm cm}^{2}\,
{\rm g}^{-1}}\right)^{-1}.
\end{equation}
Note that Wolfire \& Cassinelli (1987) argued that the dust must be depleted
by a factor of $\sim 4$ from the average interstellar value 
for the formation of massive stars.
If so, the opacity will be lower by the same factor, and  this will
affect the critical metallicity $Z_{\rm dr}$ as well.
For $Z>Z_{\rm dr}$, the infall is reversed by blowing out the outer envelope
before the luminosity reaches the Eddington limit due to electron
scattering.  In this regime, metals act against prolonged accretion.  For
$Z<Z_{\rm dr}$, the radiation force onto the outer envelope is always below
gravity, since the protostellar luminosity never exceeds the Eddington limit.
When $L_{\ast} \simeq L_{\rm Edd}$, the radiation force on the radiative
precursor and on the surface layers causes the violent expansion of the
protostar.  In this regime, metals work in favor of continuing accretion by
increasing the critical accretion rate.

\section{Discussion}
Much of the recent attention on primordial stars has focussed on the
properties of massive objects. Several authors have argued that the first
generation of cosmic structures formed mainly massive and very massive stars,
and indirect evidence for a top-heavy IMF includes chemical abundances and
supernova rates in our Galaxy, as well as star formation rates in high
redshift galaxies (e.g., Larson 1998). Also, the apparent lack or dearth of
stars in the galactic halo with metallicity less than $Z/Z_\sun\sim
10^{-4}$, coupled with the theoretical prediction of a critical metallicity
for efficient fragmentation at essentially the same value of $Z$ (Bromm et
al.  2001a) have led to the suggestion that the IMF at early times differed
substantially from its present form. Finally, the initial chemical enrichment
of our Galaxy and the abundance pattern in the intracluster medium have been
considered as compelling evidence for a generation of stars with mass
$\ga$100~M$_\sun$ (Schneider et al. 2002).  In particular, these authors
have postulated that in order to explain the apparent average minimum
metallicity of $Z/Z_\sun\sim 10^{-4}$, the first stars should have had
masses in the rather narrow interval between $140~M_\sun$ and $260~M_\sun$,
ending their lives as pair-unstable supernovae and polluting the pristine gas
with the right amount of heavy elements. Higher and lower mass stars would in
fact produce massive black holes, without releasing metals (e.g., Heger \&
Woosley 2002).

While there are good theoretical reasons to believe that massive stars were
common in the early universe, the question of their actual formation has not
been studied in detail. As we know from studies of the local universe,
forming massive stars is a rare event that requires special conditions and it
is not clear that inefficient fragmentation, lack of metals and fast mass
accretion are {\it per se} sufficient ingredients to justify the
expectations.  The discovery of an extremely metal-poor halo star of
$\sim$solar mass with an iron abundance [Fe/H]=$-$5.3, twenty times lower
than the previous limit of $Z/Z_{\sun} \approx 10^{-4}$ 
(Christlieb et al. 2002), clearly indicates that low-mass star formation in 
such extreme environments is more pervasive than otherwise believed.

The results of our calculations have shown that the final outcome of the
accretion process is a sensitive function of the mass accretion rate and of
its time variation. Notwithstanding the role of other important physical
processes that we have neglected (presence of circumstellar disks, turbulence
and magnetic fields), even in the idealized case in which star formation is
mainly determined by $\dot M_{\rm acc}$ leads to uncertain predictions on the
characteristic mass scale of the first stars. The existence of a critical
mass accretion rate $\dot M_{\rm crit}\simeq 4\times
10^{-3}~M_\sun$~yr$^{-1}$ separates solutions in which very massive
objects with mass $\gg 100~M_\sun$ can form, provided there is sufficient
material in the parent clumps, from others where the upper mass limit
decreases as $\dot M_{\rm acc}$ increases.  In the case of the time dependent
accretion rate from the numerical simulations of Abel et al. (2002),
protostellar growth can continue unimpeded by radiation forces onto the
radiative precursor.  However, slight deviations of the actual accretion rate
from the empirical form of eq.~(10) may result in a large difference in the
final mass of the formed stars.  Obviously, more refined models of the
collapse phase should be carried out to verify the robustness or uniqueness
of the behavior of the accretion rate in the high-mass regime ($M_\ast
\ga 60~M_\sun$) which is critical for prolonged accretion.  If the
decrease of $\dot M_{\rm acc}$ is slower than that given by eq.~(10),
accretion can be stopped at around a few tens to 100 $M_{\sun}$, thus not
allowing the formation of the objects which are considered responsible for the
initial enrichment of the primordial gas.

Another important result of our study is the appreciation of the sensitivity
of the evolution on the metal abundance.  In the $Z=0$ case, the value of the
actual accretion rate from eq. (1) is similar to the critical one.  For the
metal enriched case, $\dot M_{\rm acc}$ is expected to be smaller, as a
result of the reduced gas temperatures of the prestellar clumps. On the other
hand, $\dot M_{\rm crit}$ increases at higher $Z$, as shown in \S 4.  Hence,
in a slightly metal-enriched gas, $\dot M_{\rm acc}$ would be lower than
$\dot M_{\rm crit}$, leading to continued accretion and favoring the
formation of very massive stars.

Considering the mass scale for cloud fragmentation, recent studies have found
a transition from massive ($\sim 10^3 ~M_\sun$) to low-mass ($\sim 1
M_\sun$) clumps at a metallicity of $Z_{\rm frag} \sim 10^{-4} Z_{\sun}$ 
(Bromm et al. 2001a; Schneider et al. 2002). 
However, as discussed in the previous
section, there appears to be another critical metallicity.  If $Z\ga Z_{\rm
dr} \sim 0.01 Z_{\sun}$, radiation pressure onto the dust shell surrounding a
protostar becomes important and limits the amount of matter that can be
accreted. Therefore, one can argue that there are three regimes that
determine the maximum value of the stellar mass: for $Z<Z_{\rm frag}$, the
limit is determined by the the stellar lifetime and/or by the expansion of
the ionized region; for $Z_{\rm frag}<Z< Z_{\rm dr}$ by the fragmentation
scale, and by the reversal of the gas inflow due to radiation pressure
onto the dust shell for $Z_{\rm dr}<Z$. In the first case, massive stars form
predominantly, while in the other two cases their production is possible,
although not typical.

In this work, we have found that if $\dot M_{\rm acc}>\dot M_{\rm crit}$,
radiation pressure on the ionized outer layers of the protostar causes a
violent expansion of the radius.  Since we have treated protostellar
evolution as a series of hydrostatic configurations, the final outcome of
this dynamic phase could not be fully investigated.  We have suggested that
the rapid swelling leads to the ejection of the external layers, to the
onset of powerful mass loss through a wind, and eventually to the complete halt
of accretion.  Note that stellar pulsations due to the 
$\epsilon$-mechanism triggered by the onset of the CN-cycle can lead
to moderate mass loss in metal-rich stars, but in the case of metal-free 
stars the loss rate is strongly reduced (Baraffe, Heger, \& Woosley 2001).

There are, however, other possibilities that need to be considered.
For example, the stellar wind may not be strong enough to blow away all the
envelope material.  With no accretion, the central star would quickly
contract to the main sequence, decreasing substantially its mass loss
activity.  Then, the remaining circumstellar matter would be able to collapse
and recurrent accretion might be possible.  

Deviations from spherical symmetry is also crucial.  The stellar wind might
blow in a bipolar jet-like configuration, without affecting accretion from
the disk.  Even if the accretion rate through the disk is higher than the
critical rate, the protostar would adjust itself in a state marginally
below $\dot M_{\rm crit}$ by shedding the remainder material as a jet-like
wind.  Such energetic jet could disrupt the entire parent clump and 
quench the supply of matter to the disk (e.g., Nakano, Hasegawa, 
\& Norman 1995).  

When the accretion proceeds in a disk-like fashion, 
the star loses energy faster than in the case of spherical accretion 
since the stellar radiation can escape in the optically thin 
polar directions.  
The reduced entropy level in the star results in a smaller 
protostellar radius (Palla \& Stahler 1993; Hartmann, Cassen, \& Kenyon 1997). 
Thus, at the same protostellar mass, the accretion luminosity is
higher in the disk accretion than in the spherical case.
This suggests that accretion in the polar direction is halted earlier on. 
On the other hand, if the parent clump is not disrupted by 
sweeping-up of the matter in the polar direction, accretion can proceed 
through the disk since only the matter at large angles from the disk is 
subject to the radiation force (Nakano 1989).
To decide whether the disk-like configuration is more suitable or not 
for continuing accretion, 2D or 3D hydrodynamical
studies must be carried out, similar to those now available for the 
present-day case (e.g., Yorke \& Sonnhalter 2002).
Recently Tan \& McKee (2002) studied rotational effects on halting 
the protostellar accretion onto metal-free stars using a
simplified treatment. They point out the difficulty of forming very
massive objects (M>>100 Msun) because of the flow reversal in the polar
directions. However, we believe that this conclusion is uncertain since
accretion can still proceed through the circumstellar disk. 
Thus, the problem of making very massive stars is still open.

\section{Conclusions} 

We have studied the evolution of extremely low metallicity
($Z/Z_\sun=0-5 \times 10^{-3}$) protostars accreting at very fast 
rates $\dot{M}_{\rm
acc} \sim 10^{-3}-10^{-2} ~M_{\sun}$~yr$^{-1}$, as is expected to occur in
the first star forming sites of cosmological structures. The main results can
be summarized as follows:

\begin{itemize}
\item
The earliest stages of the protostellar growth
are qualitatively the same independent of the 
details of the evolution of the mass accretion rate, or of the presence
of trace abundances of heavy elements. These phases correspond to 
the adiabatic accretion, the propagation of the luminosity wave, 
and the Kelvin-Helmholtz (KH) contraction. However, the mass range covered
in each phase does depend on $\dot{M}_{\rm acc}$, and in the parameter space
explored here extends up to $\sim 60~M_{\sun}$.

\item
For more massive protostars, the evolution follows two fundamentally
different branches, depending on the instantaneous value of the mass
accretion rate. If $\dot{M}_{\rm acc}<\dot M_{\rm crit} \sim 4\times 10^{-3}
M_{\sun}$ yr$^{-1}$, protostars relax to ordinary H-burning objects while
still accreting.  In the opposite case, $\dot{M}_{\rm acc}>\dot M_{\rm crit}$,
the protostellar luminosity reaches the Eddington limit before the onset of
nuclear burning at the center.  This phase is soon followed by a rapid
expansion of the radius, possibly leading to reversal of the accretion flow.

\item
The two regimes yield upper mass limits that differ considerably: for
$\dot{M}_{\rm acc}\ga\dot M_{\rm crit}$, the maximum mass is 
$M_{\rm max}\la 300~M_\sun$ and {\it decreases} with $\dot{M}_{\rm acc}$; in
the opposite case, $M_{\rm max}\gg 300~M_\sun$, allowing the formation of
very massive objects.

\item
Under a more realistic case of time dependent accretion, such as
that proposed by Abel et al. (2002) where an initially high $\dot{M}_{\rm
acc}$ begins to decrease only at high protostellar masses ($\ga
60~M_{\sun}$), the evolution resembles the case $\dot{M}_{\rm acc}<
\dot{M}_{\rm crit}$.  Therefore, the star is expected to grow in mass until
the accretion rate drops substantially and the protostar joins the main
sequence. At this time, the expansion of an ionized region would likely stop 
further accretion ($\ga 460~M_{\sun}$; Omukai \& Inutsuka 2002). 
This limit is similar to that imposed by the amount of mass that can 
be accreted by the star during the entire main-sequence lifetime 
($600~M_{\sun}$; Abel et al. 2002).

\item
In all cases considered, deuterium burning and the $pp$-chain do not produce
enough energy to counteract the effect of Kelvin-Helmoltz contraction.  It is
only with the beginning of the CN-cycle that nuclear burning has
significant consequences for the protostellar evolution.

\item
A slight amount of metals ($Z/Z_{\sun} \sim 10^{-4} - 10^{-2}$) 
favors continuing 
accretion by increasing the value of the critical accretion rate below which
accretion can continue unimpeded. Thus, a non-zero metallicity favors the
formation of very massive protostars. Above a threshold metallicity of
$Z\simeq 0.01~Z_\sun$, the situation is reversed because of the 
feedback of radiation pressure on dust grains, as in the standard solar
metallicity case.

\item
Almost throughout the protostellar phase, the accreting envelope is optically
thick.  Although the stellar surface is as hot as $\sim 10^{5}$~K, the
photospheric temperature remains limited to $\sim$6000~K. Only after the
protostellar mass exceeds $\ga 100M_{\sun}$ and the density in the envelope
decreases substantially ($\sim 5 \times 10^{-14}{\rm g~cm^{-3}}$ 
at the photosphere), the photospheric
temperature begins to increase.  In the case of the Abel et al. evolution, the
photosphere coincides with the stellar surface for protostellar masses above
$500 M_{\sun}$.  Beyond this value, the hot stellar surface becomes optically
visible.

\item
Because of the high effective temperatures ($\sim 10^{5}$K), population III
stars have been considered an efficient source of H and He ionizing photons,
particularly in the case of star clusters consisting only of primordial
ZAMS stars (Tumlinson \& Shull 2000; Bromm et al. 2001b; Schaerer 2002).
However, during the main accretion phase, the ionizing photons are degraded to
optical photons in the precursor and cannot create a standard HII region.  If
the accretion phase is prolonged, the total emissivity of ionizing photons
might be reduced by a non-negligible fraction with important consequences on
the history of the re-ionization of the universe.

\end{itemize}

\acknowledgements The work of K.O. is supported in part by 
Research Fellowships of the Japan Society for the Promotion of Science 
for Young Scientists, grant 6819. The research of F.P.has been 
partly supported by grant COFIN 2002-MIUR to INAF--Osservatorio Astrofisico
di Arcetri. 


\newpage
\bigskip
\centerline{\bf Figure Caption}

\plotone{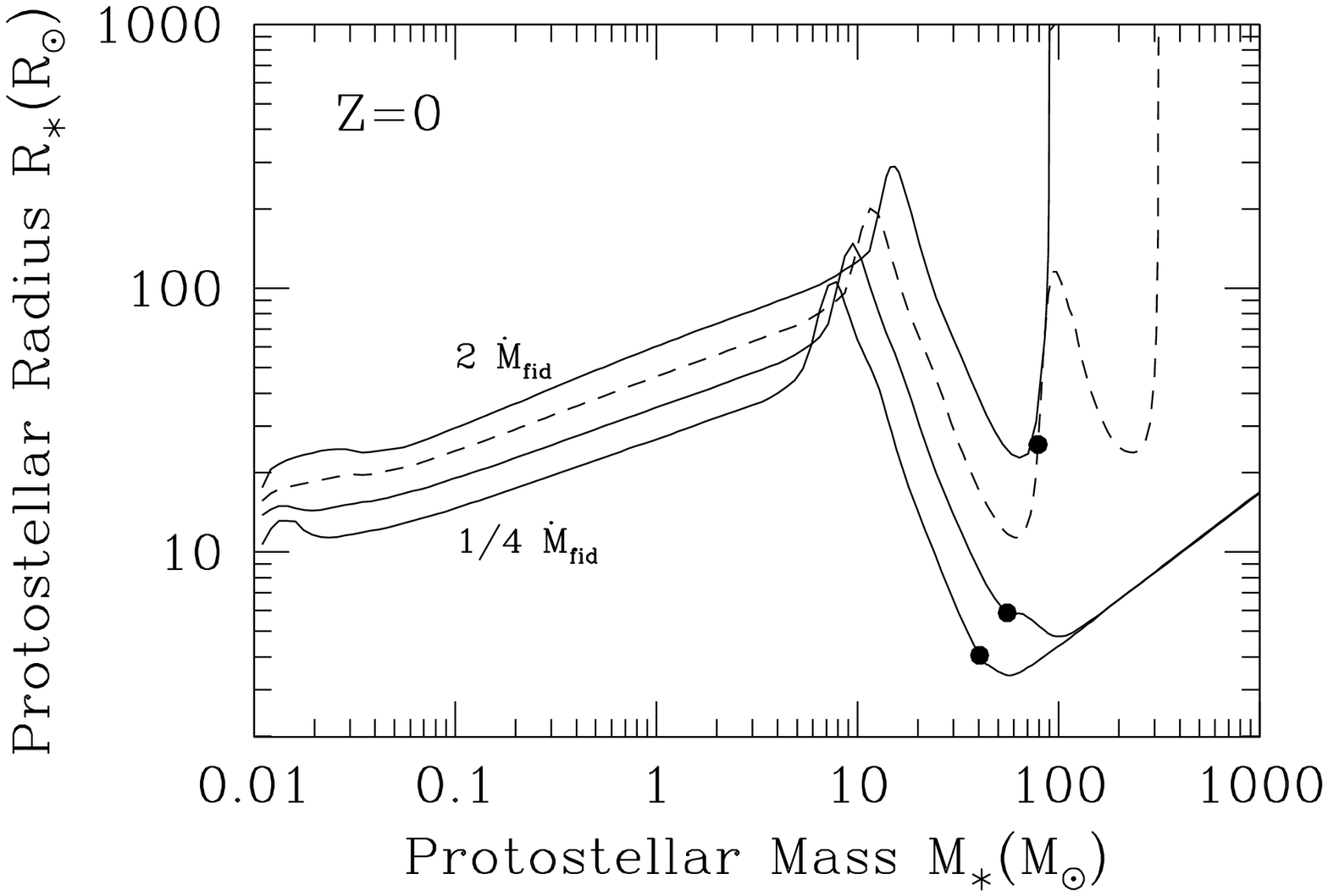}
\figcaption[f1.eps]{Mass-radius relations for metal-free
 protostars evolving with an accretion rate $\dot{M}_{\rm acc}=1/4, 1/2, 1, 
2 \dot{M}_{\rm fid}$ (from top to bottom). The dashed line is for
$\dot{M}_{\rm acc}= \dot{M}_{\rm fid}$, 
where the fiducial value is  $\dot{M}_{\rm fid}=4.4 
\times 10^{-3} M_{\sun}$ yr$^{-1}$.   
The filled circles indicate the onset of H-burning via the CN-cycle. 
\label{fig:MR_prm}}

\plotone{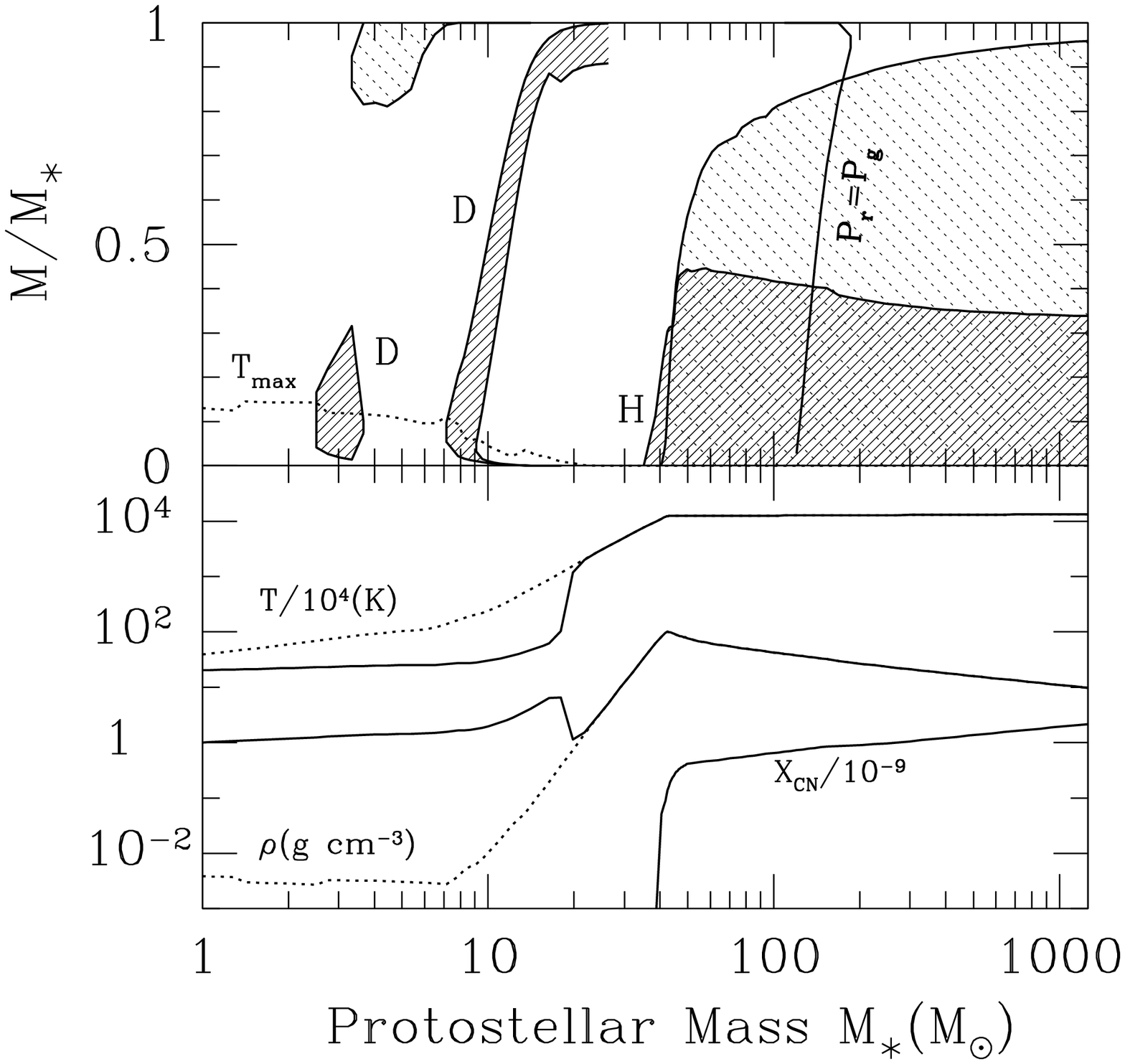}
\figcaption[f2.eps]{The internal structure of a protostar 
accreting at $\dot{M}_{\rm acc}=1/4 \dot{M}_{\rm fid}
=1.1 \times 10^{-3} M_{\sun}/{\rm yr}$ as a function of mass.
{\it Top panel}:
Evolution of the nuclear burning region and of the convective zone as a 
function of the relative mass $M/M_{\ast}$.
Regions where the energy generation by D and H burning exceed 10\% of 
the average energy generation rate $L_{\ast}/M_{\ast}$ are shown as hatched
areas.  The extent of the convective zone is displayed by the dotted area.
The solid curve labeled $P_{\rm r}=P_{\rm g}$ is the locus where the
radiation pressure equals the gas pressure.  Radiation pressure dominates to
the right of this curve.  The location of the temperature peak is also
indicated by the dotted line labeled $T_{\rm max}$.  {\it Bottom panel}:
 The evolution of temperature and density at the center (solid lines) and at
the position of the peak temperature (dotted lines).  The CN abundance in the
convective core is also shown.  Each variable is normalized to $10^{4}$~K,
1~g~cm$^{-3}$, and $10^{-9}$, respectively.  
\label{fig:core_qrt}}

\plotone{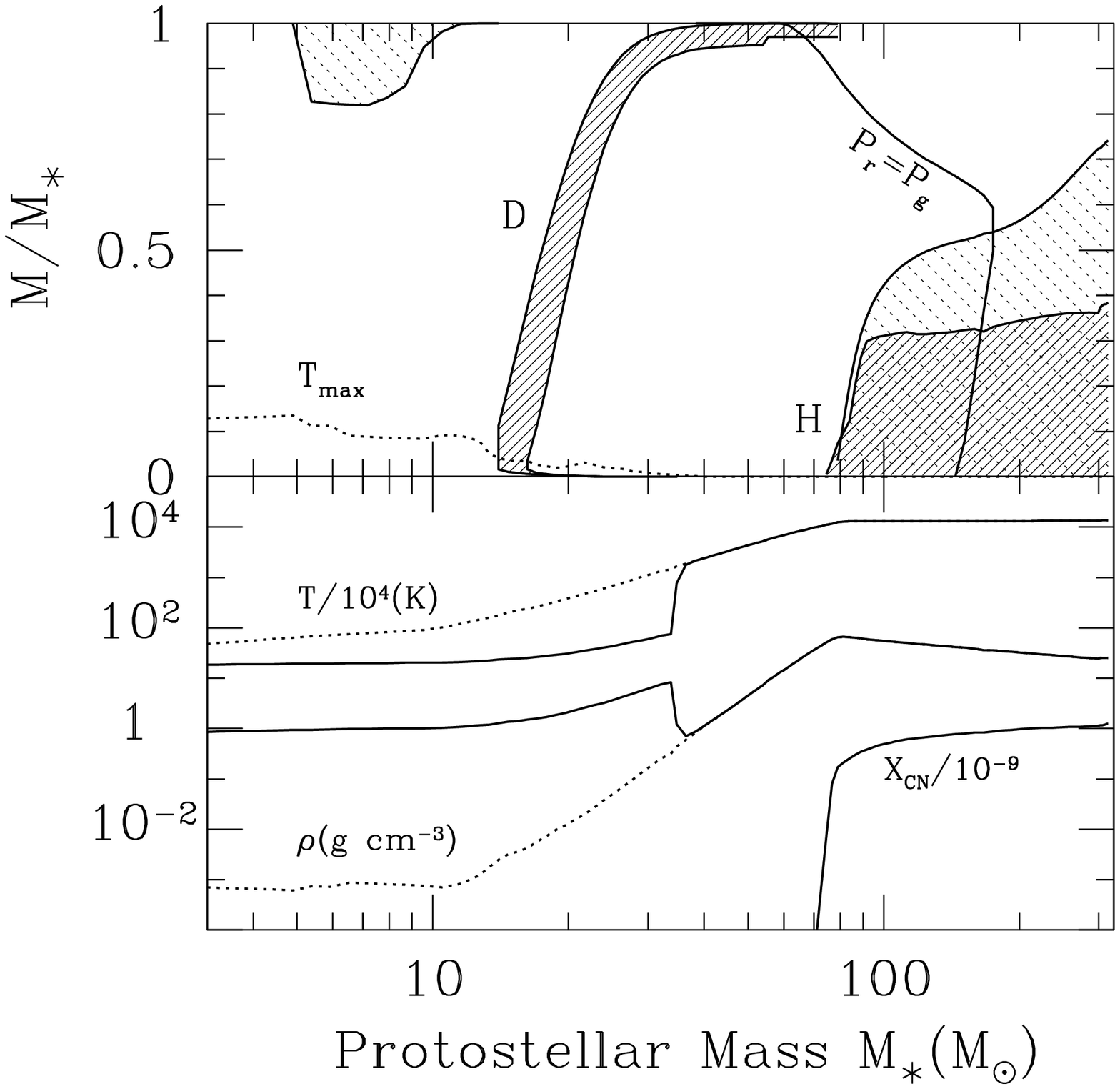}
\figcaption[f3.eps]{Same as Fig.~2,
but for $\dot{M}_{\rm acc}=\dot{M}_{\rm fid}=4.4 \times 10^{-3} M_{\sun}/{\rm yr}$.
\label{fig:core_fid}}

\plotone{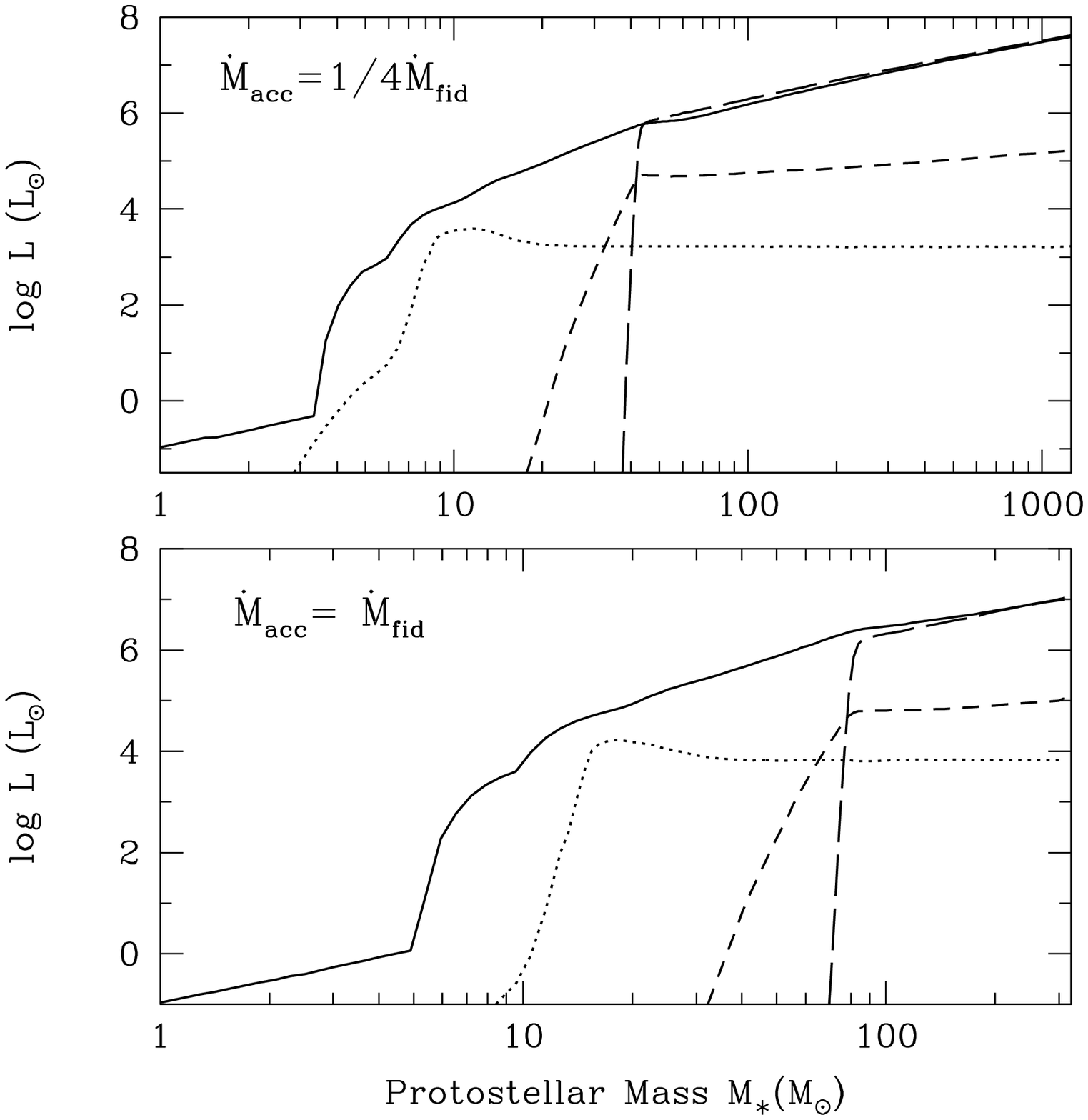}
\figcaption[f4.eps]{The evolution of the interior luminosity (solid)
 and the contribution due to deuterium burning (dotted), the {\it p-p} 
 chain (short-dashed), and the CN-cycle (long-dashed) 
for $\dot{M}_{\rm acc}=1/4 \dot{M}_{\rm fid}=1.1 \times 10^{-3} 
M_{\sun}/{\rm yr}$ ({\it top panel}) and 
for $\dot{M}_{\rm acc}= \dot{M}_{\rm fid}=4.4 \times 10^{-3} 
M_{\sun}/{\rm yr}$ ({\it bottom panel}).
\label{fig:Lnuc}}

\plotone{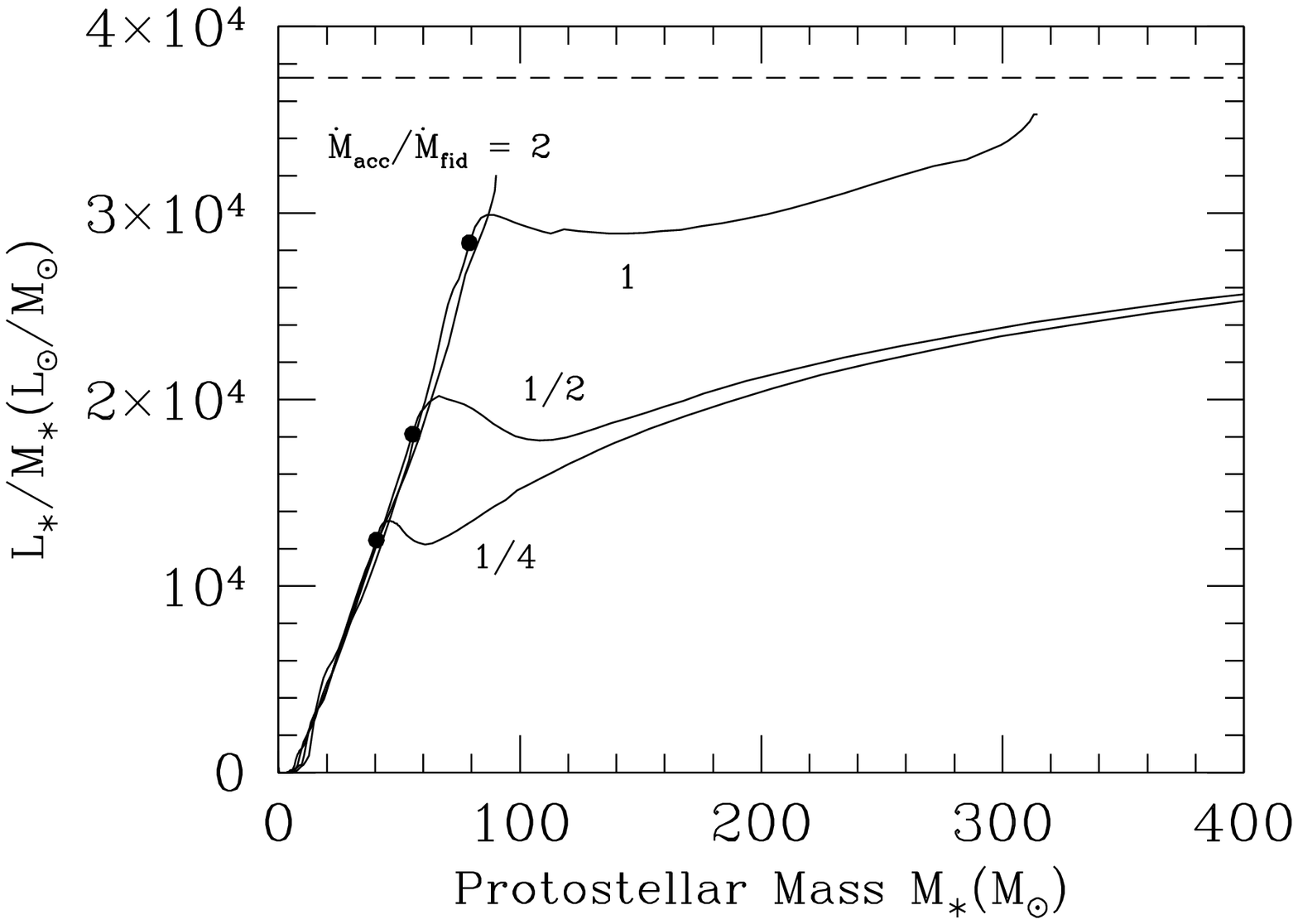}
\figcaption[f5.eps]{Evolution of the interior luminosity-to-mass ratio 
as a function of mass for different values of the ratio 
$\dot{M}_{\rm acc}/\dot{M}_{\rm fid}$.
The dashed horizontal line corresponds to the Eddington limit due to 
electron scattering.
\label{fig:ML}}

\plotone{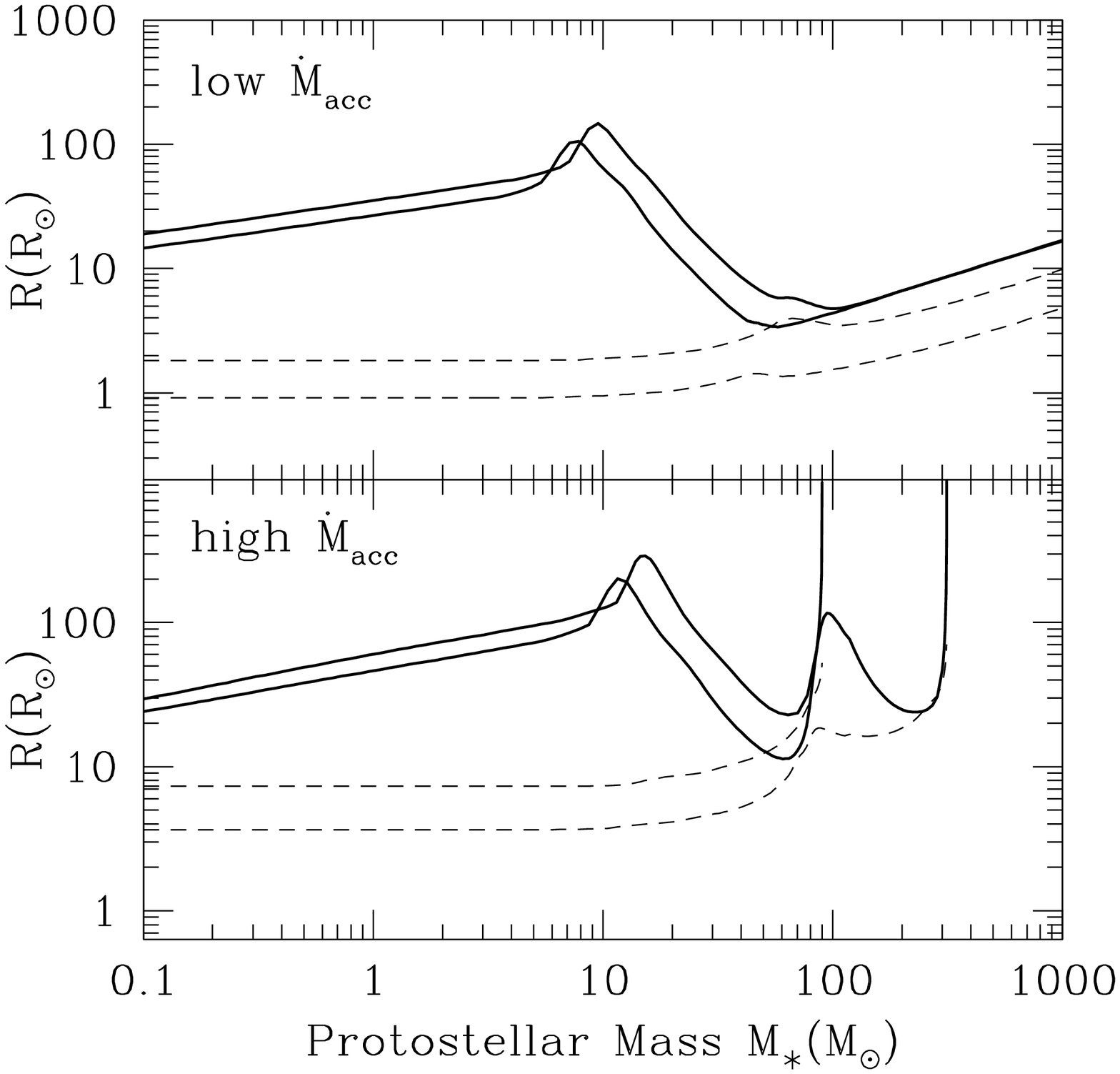}
\figcaption[f6.eps]{Evolution of the core radius (solid) and the Eddington 
radius (dashed) for values of the accretion rate lower ({\it top panel}) and
higher ({\it bottom panel}) than the fiducial one.
\label{fig:MRedd}}

\plotone{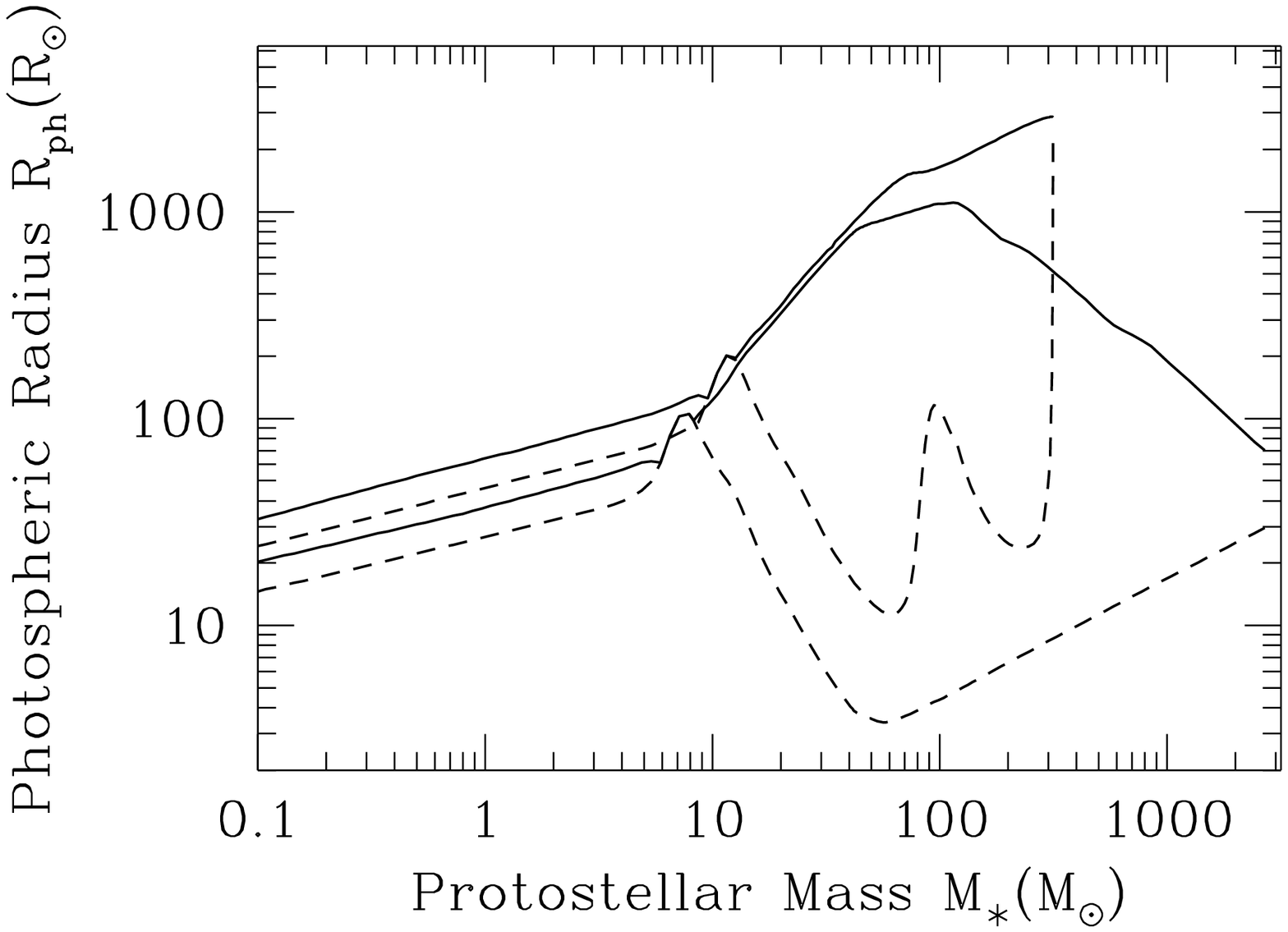}
\figcaption[f7.eps]{The evolution of the photospheric radius
for two cases with $1/4 \dot{M}_{\rm fid}$ and $\dot{M}_{\rm fid}$,
respectively. The dashed lines represent the evolution of the core radius. 
\label{fig:MRph}}

\plotone{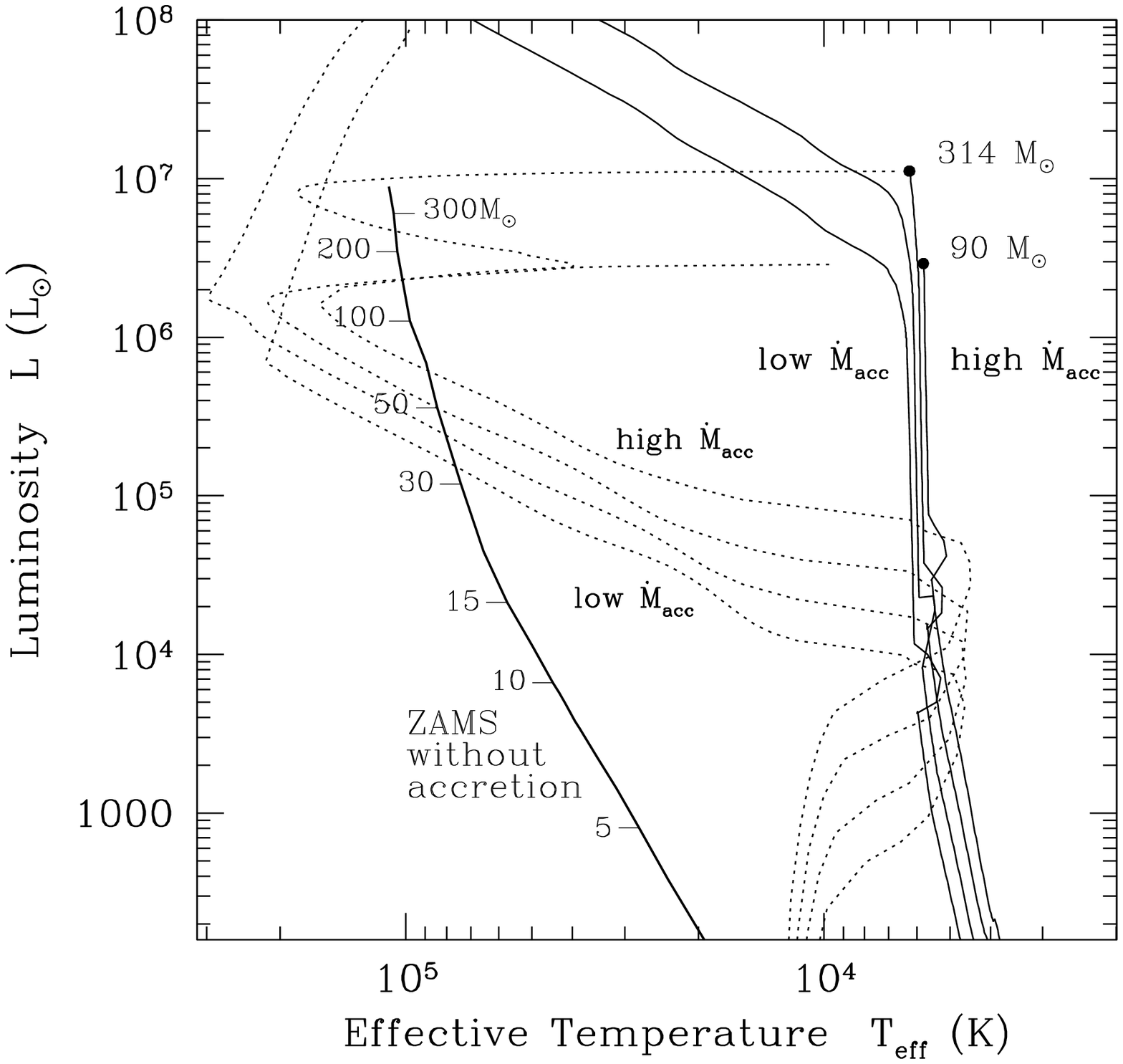}
\figcaption[f8.eps]{HR diagram for primordial protostars.
For comparison, we also show the locus of the metal-free ZAMS stars
as computed by Marigo et al. (2001) for $M_{\ast}< 100 M_{\sun}$ and by
Bromm et al. (2001b) for higher masses.
The dotted lines show the temperature and luminosity at the stellar surface.
\label{fig:HR}}

\plotone{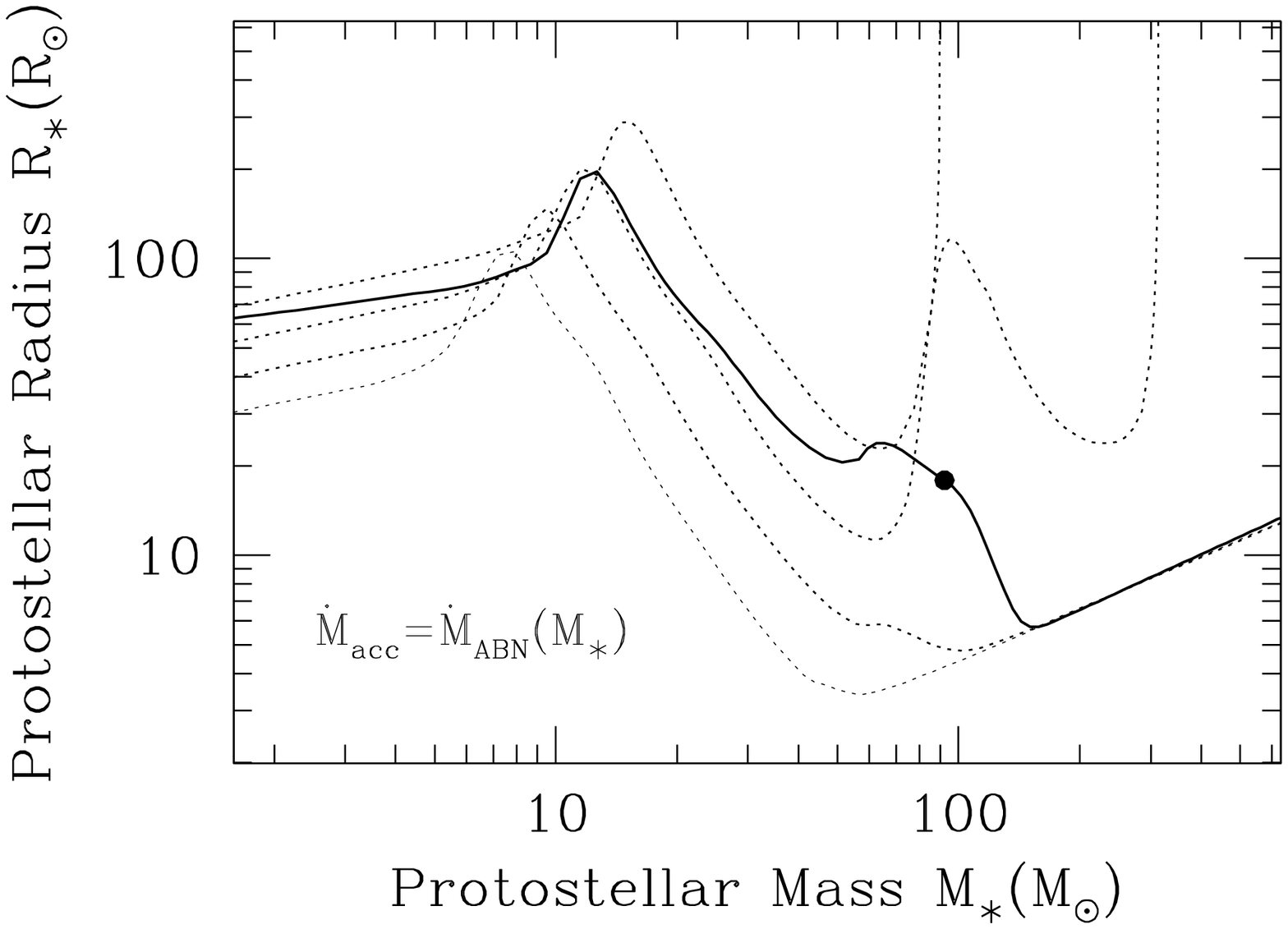}
\figcaption[f9.eps]{The evolution of the protostellar radius 
with the time dependent accretion rate of Abel et al. (2002).
The filled circle marks the onset of H-burning.
The thin dashed lines represent the evolution for constant 
$\dot{M}_{\rm acc}=1/4,1/2,1,$ and $2\dot{M}_{\rm crit}$. 
\label{fig:MR_abn}}

\plotone{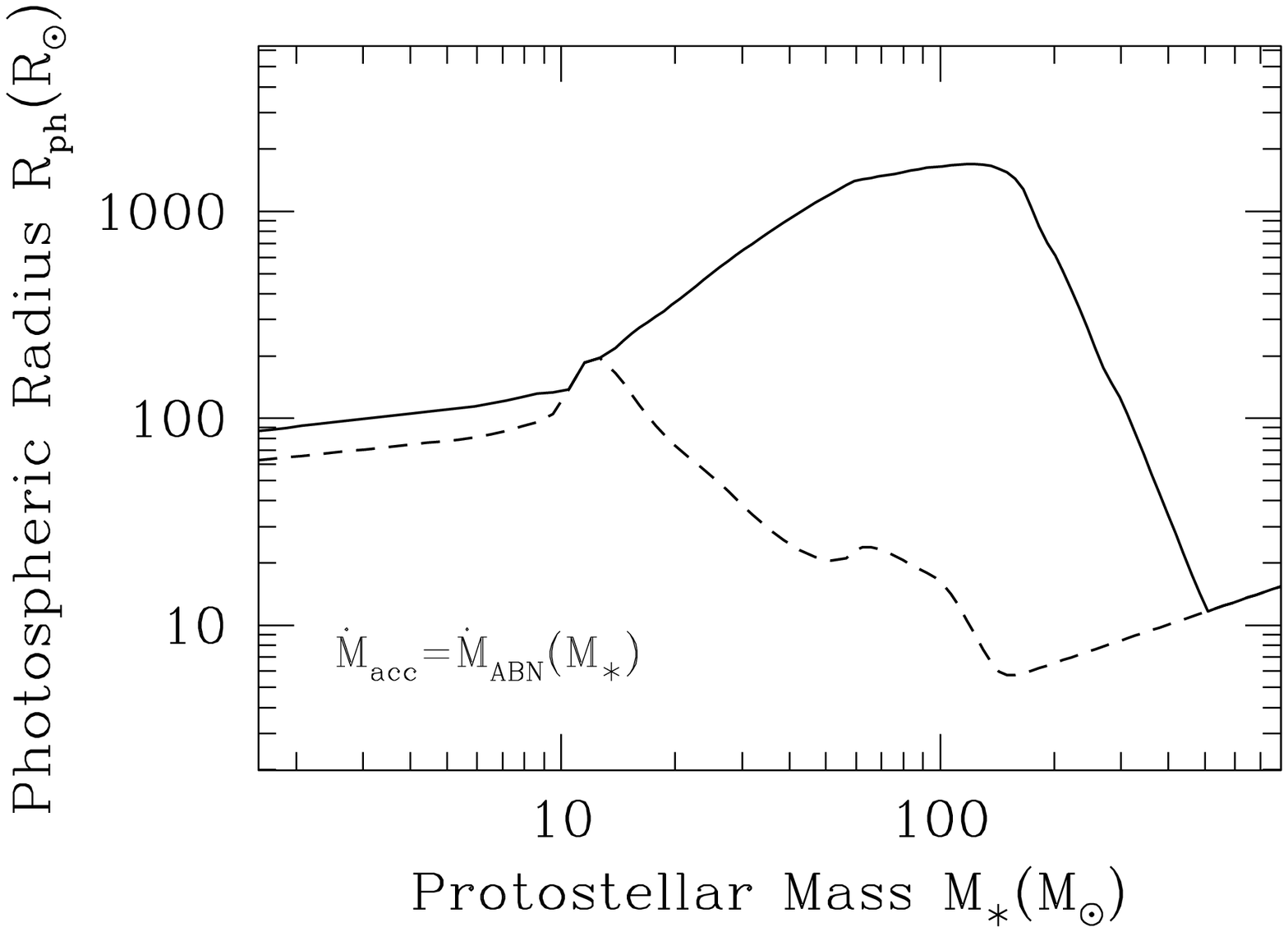}
\figcaption[f10.eps]{Run of the photospheric radius in the time 
dependent accretion rate models. The protostellar radius is shown 
by the dashed line. 
\label{fig:MR_abn2}}

\plotone{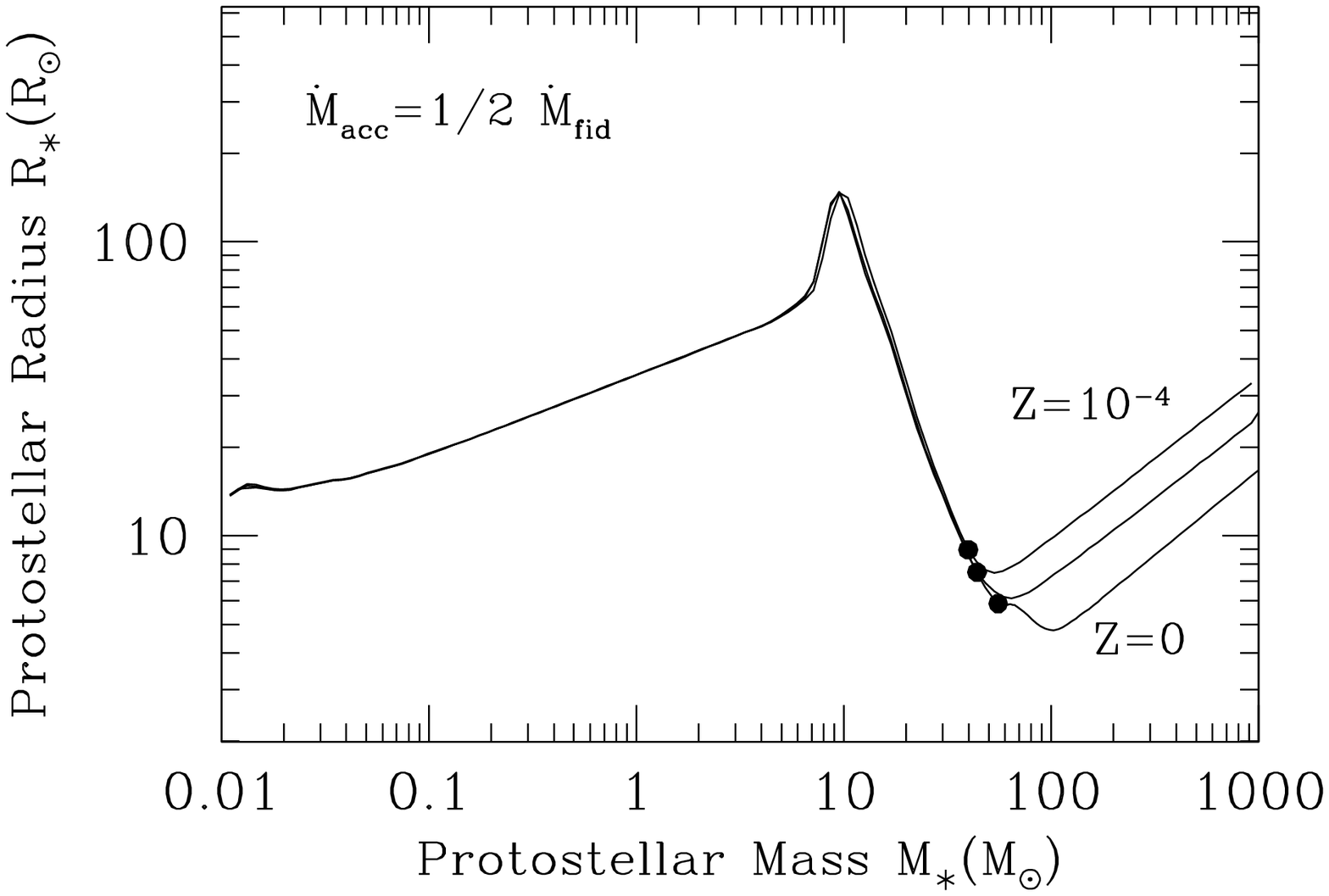}
\figcaption[f11.eps]{The protostellar mass-radius relations 
at fixed $\dot{M}_{\rm acc}=1/2 \dot{M}_{\rm fid}$ for 
different metallicities $Z=0,10^{-6}(=5 \times 10^{-5}Z_{\sun}),
10^{-4}(=5 \times 10^{-3}Z_{\sun})$. The filled circles mark the
beginning of the CN-cycle.
\label{fig:MR_half}}

\plotone{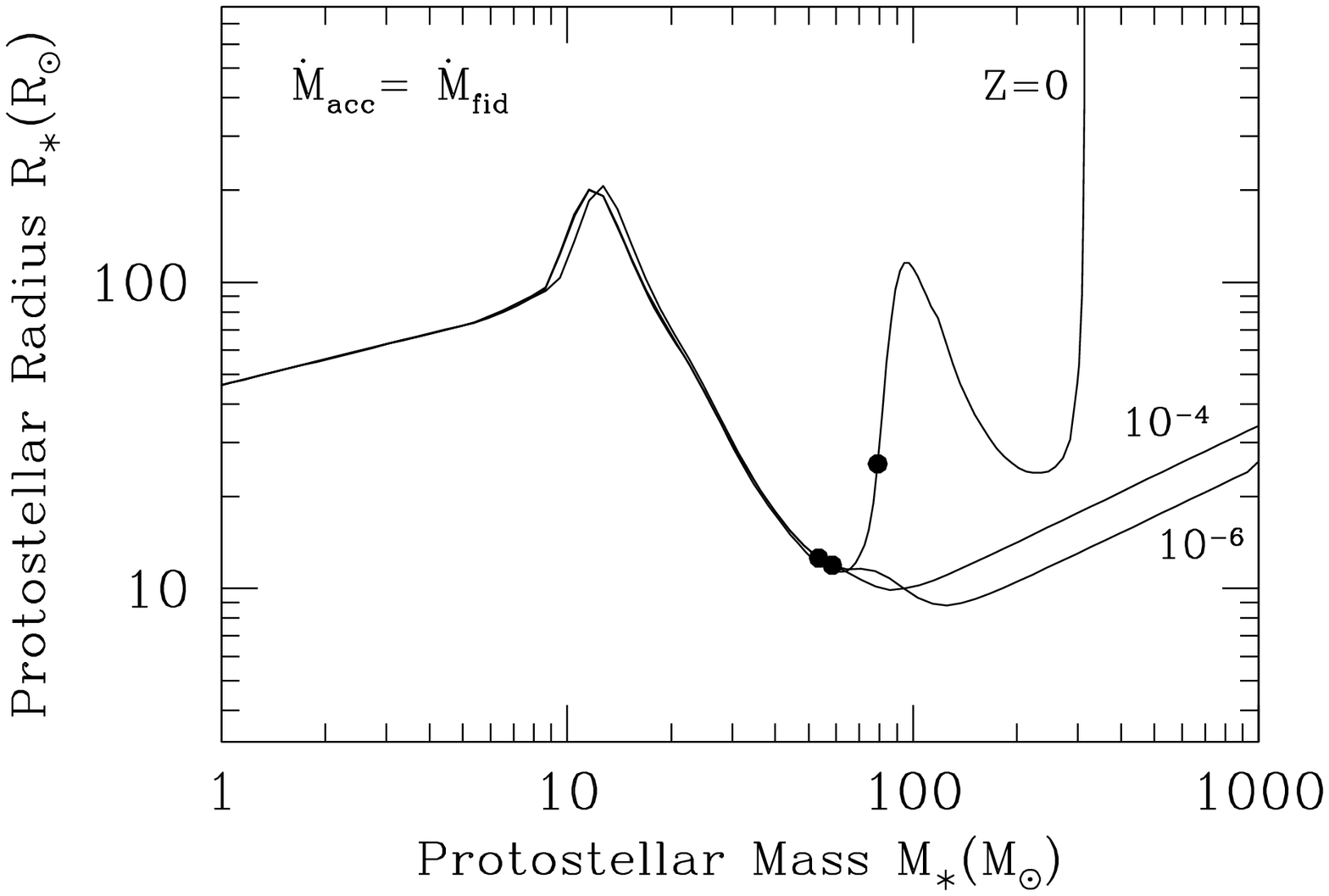}
\figcaption[f12.eps]{The protostellar mass-radius relations
computed for $\dot{M}_{\rm acc}\equiv\dot{M}_{\rm fid}=
4.4 \times 10^{-3} M_\sun$ yr$^{-1}$ at different metallicities.
\label{fig:MR_fid}}


\begin{thebibliography}{} 
\bibitem[]{} Abel, T., Bryan, G. L., \& Norman, M. L. 2000, \apj, 540, 39 
\bibitem[]{} Abel, T., Bryan, G. L., \& Norman, M. L. 2002, Science, 295, 93
\bibitem[]{} Alexander, D. R., \& Ferguson, J. W. 1994, \apj, 437, 879 
\bibitem[]{} Baraffe, I., Heger, A., \& Woosley, S. E. 2001, \apj, 550, 890 
\bibitem[]{} Beech, M., \& Mitalas, R. 1994, \apjs, 95, 517
\bibitem[]{} Bromm, V., Coppi, P. S., \& Larson, R. B. 1999, \apj, 527, L5
\bibitem[]{} Bromm, V., Coppi, P. S., \& Larson, R. B. 2002, \apj, 564, 23
\bibitem[]{} Bromm, V., Ferrara, A., Coppi, P. S., \& Larson, R. B. 2001a, 
\mnras, 328, 969
\bibitem[]{} Bromm, V., Kudritzki, R. P., \& Loeb, A. 2001b, \apj, 552, 464
\bibitem[]{} Christlieb, N., Bessell, M.S., Beers, T.C., Gustaffson, B., 
Korn, A., Barklem, P.S., Karlsson, M., Mizuno-Wiedner, M., \& Rossi, S. 
2002, Nature, 419, 904
\bibitem[]{} Fall, S. M., \& Rees, M. J. 1985, \apj, 298, 18 
\bibitem[]{} Foster, P. N., \& Chevalier, R. A. 1993, \apj, 416, 303
\bibitem[]{} Fuller, T. M., \& Couchman, H. M. P. 2000, \apj, 544, 6
\bibitem[]{} Galli, D., Palla, F., Ferrini, F., \& Penco, U. 1995, \apj, 443, 
536 
\bibitem[]{} Haiman, Z., Thoul, A. A., \& Loeb, A. 1996, \apj, 464, 523
\bibitem[]{} Hartmann, L., Cassen, P., \& Kenyon, S. J. 1997, \apj, 475, 770
\bibitem[]{} Heger, A., \& Woosley, S. E. 2002, \apj, 567, 532 
\bibitem[]{} Iglesias, C. A., \& Rogers, F. J. 1996, \apj, 464, 943
\bibitem[]{} Izotov, Yu. I., \& Thuan, T. X. 1998, \apj, 500, 188
\bibitem[]{} Kahn, F. D. 1974, \aap, 37, 149 
\bibitem[]{} Klessen, R. 2001, \apj, 550, L77
\bibitem[]{} Larson, R. B. 1969, \mnras, 145, 271
\bibitem[]{} Larson, R. B. 1998, \mnras, 301, 569
\bibitem[]{} Larson, R. B., \& Starrfield, S. 1971, \aap, 13, 190 
\bibitem[]{} Lenzuni, P., Chernoff, D. F., \& Salpeter, E. E. 1991,
\apjs, 76, 759
\bibitem[]{} Marigo, P., Girardi, L., Chiosi, C., \& Wood, P.
2001, \aap, 371, 152 
\bibitem[]{} Nakamura, F., \& Umemura, M. 1999, \apj, 515, 239 
\bibitem[]{} Nakamura, F., \& Umemura, M. 2001, \apj, 548, 19 
\bibitem[]{} Nakamura, F., \& Umemura, M. 2002, \apj, 569, 549 
\bibitem[]{} Nakano, T. 1989, \apj, 345, 464
\bibitem[]{} Nakano, T., Hasegawa, T., \& Norman, C.  1995, \apj, 450, 183
\bibitem[]{} Omukai, K. 2000, \apj, 534, 809
\bibitem[]{} Omukai, K. 2001, \apj, 546, 635
\bibitem[]{} Omukai, K., \& Inutsuka, S. 2002, \mnras, 332, 59 
\bibitem[]{} Omukai, K., \& Nishi, R. 1998, \apj, 508, 141 
\bibitem[]{} Omukai, K., \& Palla, F. 2001, \apj, 561, L55 (Paper I) 
\bibitem[]{} Palla, F., Salpeter, E. E., \& Stahler, S. W. 1983, \apj, 271, 632
\bibitem[]{} Palla, F., \& Stahler, S. W. 1991, \apj, 375, 288
\bibitem[]{} Palla, F., \& Stahler, S. W. 1993, \apj, 418, 414
\bibitem[]{} Schaerer, D. 2002, \aap, 382, 28 
\bibitem[]{} Schneider, R., Ferrara, A., Natarajan, P., \& Omukai, K. 2002, 
\apj, 571, 30
\bibitem[]{} Stahler, S. W., Palla, F., Salpeter, E. E. 1986, \apj, 302, 590 
(SPS) 
\bibitem[]{} Stahler, S. W., Shu, F. H., \& Taam, R. E. 1980, \apj, 241, 637 
\bibitem[]{} Tan, J. C., \& McKee, C. F. 2002, in proceedings of the 13th 
Annual Astrophysics Conference in Maryland: The Emergence of Cosmic Structure,
 eds. S. Holt and C. Reynolds, (AIP), (astro-ph/0212283)
\bibitem[]{} Tegmark, M., Silk, J., Rees, M. J., Blanchard, A., Abel, T., 
\& Palla, F. 1997, \apj, 474, 1 
\bibitem[]{} Tomisaka, K. 1996, \pasj, 48, L97 
\bibitem[]{} Tsuribe, T., \& Inutsuka, S. 2001, \apss, 276, 1097
\bibitem[]{} Tumlinson, J., \& Shull, J. M., 2000, \apj, 528, L65
\bibitem[]{} Uehara, H., \& Inutsuka, S. 2000, \apj, 531, L91
\bibitem[]{} Uehara, H., Susa, H., Nishi, R., Yamada, M., \& Nakamura, T.
1996, \apj, 473, L95
\bibitem[]{} Yorke, H.W., \& Sonnhalter, C. 2002, \apj, 569, 846
\bibitem[]{} Yoshii, Y., \& Saio, H. 1986, \apj, 301, 587
\bibitem[]{} Wolfire, M. G., \& Cassinelli, J. P. 1987, \apj, 319, 850
\end{thebibliography}
\end{document}